\documentclass[preprint,preprintnumbers,aps,superscriptaddress,nofootinbib,tightenlines,floatfix]{revtex4}

\usepackage{amsmath,amssymb}
\usepackage{graphicx}
\usepackage{bm}
\usepackage{comment,color,ulem}

\newcommand{\beq}{\begin{equation}}
\newcommand{\eeq}{\end{equation}}
\newcommand{\bea}{\begin{eqnarray}}
\newcommand{\eea}{\end{eqnarray}}

\newcommand{\gsim}{\raisebox{-0.7ex}{$\stackrel{\textstyle >}{\sim}$ }}

\newcommand{\coarse}{$b\approx0.125$~fm }
\newcommand{\fine}{$b\approx0.09$~fm }

\def\OMIT#1{{}}

\def\d{{\delta}}
\def\D{{\Delta}}

\def\g{{\gamma}}
\def\G{{\Gamma}}

\def\S{{\Sigma}}

\def\tr{\text{tr}}

\def\mc#1{{\mathcal #1}}

\def\eqref#1{{(\ref{#1})}}

\def\DFV{{\Delta^{(FV)}}}
\newcommand{\lthree}{4.04(40)({}^{73}_{55})}
\newcommand{\lfour}{4.30(51)({}^{84}_{60})}
\newcommand{\lthreeRes}{17(5)({}^{\phantom{1}6}_{10})}
\newcommand{\lfourRes}{0(11)(12)}
\newcommand{\fpi}{128.2(3.6)({}^{4.4}_{6.0})({}^{1.2}_{3.3})}
\newcommand{\fpif}{1.062(26)({}^{42}_{40})}
\newcommand{\rone}{0.306(9)({}^{10}_{14})}

{\count255=\time\divide\count255 by 60 \xdef\hourmin{\number\count255}
  \multiply\count255 by-60\advance\count255 by\time
  \xdef\hourmin{\hourmin:\ifnum\count255<10 0\fi\the\count255}}

\begin{document}

\begin{figure}[!t]

  \vskip -1.5cm
  \leftline{\includegraphics[width=0.25\textwidth]{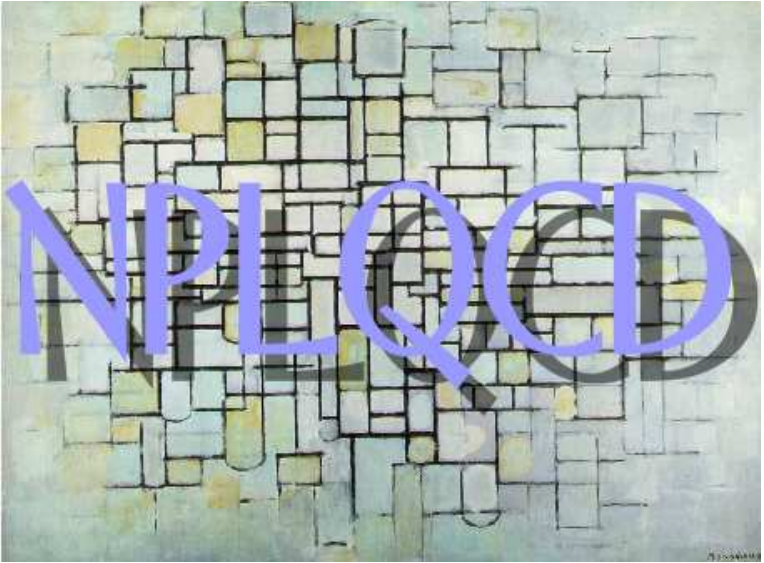}}
\end{figure}

\preprint{\vbox{ 
	\hbox{JLAB-THY-11-1380}
    \hbox{NT-LBNL-11-013}
    \hbox{NT@UW-11-14}
    \hbox{UCB-NPAT-11-008}
	\hbox{UNH-11-4} 
    }}

\title{$SU(2)$ Low-Energy Constants from Mixed-Action Lattice QCD}


\author{S.R.~Beane} 
\affiliation{Albert Einstein Zentrum f\"ur Fundamentale Physik,
Institut f\"ur Theoretische Physik,
Sidlerstrasse 5,
CH-3012 Bern, Switzerland}
\affiliation{Department of Physics, University
  of New Hampshire, Durham, NH 03824-3568, USA}
\author{W.~Detmold}
\affiliation{Department of Physics, College of William and Mary,
  Williamsburg, VA 23187-8795.}  \affiliation{Jefferson Laboratory,
  12000 Jefferson Avenue, Newport News, VA 23606.}  
 \author{P.M.~Junnarkar}
\affiliation{Department of Physics, University
  of New Hampshire, Durham, NH 03824-3568, USA}  
\author{T.C.~Luu} \affiliation{N
  Division, Lawrence Livermore National Laboratory, Livermore, CA
  94551.}  
\author{K.~Orginos} \affiliation{Department of Physics,
  College of William and Mary, Williamsburg, VA 23187-8795.}
\affiliation{Jefferson Laboratory, 12000 Jefferson Avenue, Newport
  News, VA 23606.} 
\author{A.~Parre\~no}
\affiliation{Departament d'Estructura i Constituents de la Mat\`{e}ria and
Institut de Ci\`encies del Cosmos, 
Universitat de Barcelona,  E--08028 Barcelona, Catalunya.}
\author{M.J.~Savage}
\affiliation{Department of Physics, University of Washington, Seattle,
  WA 98195-1560.} 
\author{A.~Torok} \affiliation{Department of
  Physics, Indiana University, Bloomington, IN 47405.}
\author{A.~Walker-Loud} 
  \affiliation{Nuclear Science Division, Lawrence Berkeley National Laboratory, Berkeley, CA 94720.}
  \affiliation{Department of Physics, University of California, Berkeley, CA 94720, USA}
\collaboration{ NPLQCD Collaboration } \noaffiliation \vphantom{}


\begin{abstract}
  An analysis of the pion mass and pion decay
  constant is performed using 
  mixed-action Lattice QCD calculations with
  domain-wall valence quarks on ensembles of rooted, staggered $n_f =
  2+1$ configurations generated by MILC.  
Calculations were performed
  at two lattice spacings of $b \approx 0.125$~fm and $b \approx 0.09$~fm, at
  two strange quark-masses, multiple light quark-masses, and a number of
  lattice volumes.
The ratios of light quark to strange quark-masses are in the range $0.1 \leq m_l
/ m_s \leq 0.6$,
while pion masses are in the range $235 \lesssim m_\pi \lesssim 680$~MeV.
A two-flavor chiral perturbation theory analysis of the Lattice QCD
calculations constrains  the Gasser-Leutwyler coefficients  $\bar{l}_3$ and
$\bar{l}_4$ to be $\bar{l}_3 = \lthree$
and  $\bar{l}_4 = \lfour$.
All systematic effects in
the calculations are explored, including those from the finite
lattice space-time volume, the finite lattice-spacing, and the
finite fifth dimension in the domain-wall quark
action.  
A consistency is demonstrated between a chiral
perturbation theory analysis at fixed lattice-spacing combined with a leading order continuum
extrapolation, and the mixed-action chiral perturbation theory
analysis which explicitly includes the leading order discretization
effects. 
Chiral corrections to the pion decay constant are found to give 
$f_\pi / f = \fpif$ 
where $f$ is the decay constant in the chiral-limit,
and when combined with the experimental determination of $f_\pi$ results in a value of $f = 122.8(3.0)({}^{4.6}_{4.8})$~MeV.
The most recent scale setting by the MILC
Collaboration yields a postdiction of 
$f_\pi = \fpi$~MeV 
at the physical pion mass.
A detailed error analysis indicates that precise calculations at lighter pion masses is the single most important systematic to address to improve upon the present work.
\end{abstract}

\maketitle

\tableofcontents
\newpage

%
%
\section{Introduction \label{sec:Intro} }

\noindent 
The masses and decay constants of the pseudo-Goldstone bosons are
hadronic observables that Lattice QCD can now calculate with
percent-level accuracy in the absence of isospin breaking and
electromagnetism.  This is primarily due to the fact that the
signal-to-noise ratio of the ground state contribution to pion
correlation functions does not degrade exponentially with time.  While
Lattice QCD calculations are still being carried out at unphysically
large quark-masses, with relatively coarse lattice spacings, and in
modest volumes, chiral perturbation theory ($\chi$PT) can be
used to describe the dependence of the pseudo-Goldstone boson masses
and decay constants on these variables.  Such a description involves a
set of low-energy constants (LECs), which can be determined from
experimental measurements, or from the Lattice QCD calculations
themselves.  The LECs that are extracted from the pseudo-Goldstone
boson observables also appear in other physical processes, and
therefore accurate Lattice QCD calculations of pion and kaon
correlation functions are beginning to translate into predictive power
for other --more complicated-- observables involving pions and kaons.

$\chi$PT, the low-energy effective field theory (EFT) of QCD, provides
a systematic description of low-energy processes involving the
pseudo-Goldstone bosons~\cite{Weinberg:1978kz}.  The theory consists
of an infinite series of operators (and their coefficients, the LECs) whose
forms are constrained by the global symmetries of QCD.  The
quantitative relevance of these operators is dictated by an expansion
in terms of the pion momentum and light quark-masses suppressed by the
chiral symmetry breaking scale, $\Lambda_\chi$.  At leading order (LO)
in the two-flavor ($n_f=2$) chiral expansion, the two coefficients
that appear are determined by the pion mass, $m_\pi$, and the pion decay
constant, $f_\pi$.  At next-to-leading order (NLO), there are four new
operators in the isospin limit whose coefficients are not constrained
by global symmetries~\cite{Gasser:1983yg}; these LECs 
are the Gasser-Leutwyler coefficients.  Two of
these LECs, $\bar{l}_1$ and $\bar{l}_2$, can be reliably determined
from low-energy $\pi\pi$ scattering~\cite{Colangelo:2001df}.  
The LEC
$\bar{l}_3$ governs the size of the NLO contributions to $m_\pi$, while
$\bar{l}_4$ controls the size of the NLO contributions to $f_\pi$.
Lattice QCD, the numerical solution of QCD, provides a way
to constrain these coefficients, 
including those that depend upon the light quark-masses.
Further, as Lattice QCD calculations can be
performed to arbitrary precision with appropriate computational
resources, 
they will likely provide more precise determinations of
the LECs than can be extracted from experimental data.  
A number of lattice collaborations have recently
determined $\bar{l}_3$ and $\bar{l}_4$ using $n_f=2$, $n_f = 2+1$ and $n_f=2+1+1$
calculations of $m_\pi$ and $f_\pi$ with a variety of lattice
discretizations~\cite{Allton:2008pn,Noaki:2008iy,Aoki:2008sm,Baron:2009wt,Baron:2010bv,Aoki:2010dy,Bazavov:2010yq,Bazavov:2010hj}.
These efforts have been compiled into a review article~\cite{Colangelo:2010et} which has
performed averages of these various computational
efforts. It should be noted that there is an
increasing number of Lattice QCD calculations performed at or near the
physical
point~\cite{Aoki:2008sm,Durr:2008zz,Aoki:2009ix,Durr:2010vn,Durr:2010aw},
and it will be exciting to have reliable predictions of hadronic
observables that do not rely on $\chi$PT. 

In this work, we focus on the determination of $\bar{l}_3$ and
$\bar{l}_4$ from the pion mass and the pion decay constant using a
mixed-action (MA) calculation with domain-wall valence quarks on 
gauge-field configurations
generated with rooted, staggered sea-quarks.  
This serves to strengthen
the case that the systematic effects arising from the finite
lattice-spacing, which are unique to a given lattice discretization,
can be systematically eliminated to produce results that are
independent of the fermion and gauge lattice actions.
There are already preliminary results from mixed-action calculations which can be
found in Ref.~\cite{Aubin:2008ie}.

Section~\ref{sec:Latt} describes the details of the Lattice QCD
calculation.  In Sec.~\ref{sec:Sys}, details of the systematic
uncertainties are presented.  
Continuum and chiral extrapolations of
the results of the Lattice QCD calculations 
are detailed in Sec.~\ref{sec:ChiPT}.
Conclusions are presented in Sec.~\ref{sec:Results}.

%
%
\section{Details of the Lattice Calculation and Numerical Data\label{sec:Latt} }
\noindent
The present  work is part of a program of mixed-action lattice QCD
calculations performed 
by the NPLQCD
collaboration~\cite{Beane:2005rj,Beane:2006mx,Beane:2006pt,Beane:2006fk,Beane:2006kx,Beane:2006gj,Beane:2006gf,Beane:2007xs,Beane:2007uh,Beane:2007es,Detmold:2008fn,Beane:2008dv,Detmold:2008yn,Detmold:2008bw,Torok:2009dg}.
The strategy, initiated  by the LHP Collaboration~\cite{Renner:2004ck,Edwards:2005kw,Edwards:2005ym,Hagler:2007xi,WalkerLoud:2008bp,Bratt:2010jn}, 
is to compute domain-wall fermion~\cite{Kaplan:1992bt,Shamir:1992im,Shamir:1993zy,Shamir:1998ww,Furman:1994ky}
propagators generated on the $n_f = 2+1$ asqtad-improved~\cite{Orginos:1998ue,Orginos:1999cr}
rooted, staggered sea-quark configurations generated by the MILC Collaboration~\cite{Bernard:2001av,Bazavov:2009bb},
(with hypercubic-smeared~\cite{Hasenfratz:2001hp,DeGrand:2002vu,DeGrand:2003sf,Durr:2004as}
gauge links to improve the chiral symmetry properties of the domain-wall propagators).
The predominant reason for the success of this program
is the good chiral symmetry properties of
the domain-wall action, which significantly suppresses chiral 
symmetry breaking from the staggered sea fermions and 
discretization effects~\cite{Chen:2005ab,Chen:2006wf,Chen:2007ug}.
This particular mixed-action approach has been used to perform a detailed study
of the meson  and baryon spectrum~\cite{WalkerLoud:2008bp} 
including a comparison with predictions from the large-$N_c$ limit of QCD and 
$SU(3)$ chiral symmetry~\cite{Jenkins:1995td,Jenkins:2009wv}.  
The static and charmed baryon spectrum
were respectively  determined in Refs.~\cite{Lin:2009rx,Liu:2009jc}; 
the first calculation of the hyperon axial charges was 
performed in Ref.~\cite{Lin:2007ap};
the first calculation of the strong isospin
breaking contribution to the neutron-proton mass difference was calculated in Ref.~\cite{Beane:2006fk},
and the hyperon electromagnetic form factors were explored in Ref.~\cite{Lin:2008mr}.
The majority of calculations using this mixed-action strategy have been
performed at only one
lattice-spacing, the \textit{coarse} lattice-spacing of $b\approx 0.125$~fm;
a notable exception 
was the calculation of $B_K$~\cite{Aubin:2009jh}, which included the \textit{fine} MILC ensembles 
with $b\approx 0.09$~fm.
In Ref.~\cite{Aubin:2008wk}, very nice agreement was found between the
prediction of the  scalar $a_0$ correlation function 
from mixed-action $\chi$PT (MA$\chi$PT)
and the Lattice QCD calculations of the same correlation function~\cite{Prelovsek:2005rf}.
This was an important check of the 
understanding of
unitarity violations that are inherent in
mixed-action calculations.

%
%
\subsection{Lattice QCD Parameters\label{sec:Latt_Params} }
\noindent
In our previous
works~\cite{Beane:2005rj,Beane:2006mx,Beane:2006pt,Beane:2006fk,Beane:2006kx,Beane:2006gj,Beane:2006gf,Beane:2007xs,Beane:2007uh,Beane:2007es,Detmold:2008fn,Beane:2008dv,Detmold:2008yn,Detmold:2008bw,Torok:2009dg},
on the \coarse ensembles, domain-wall valence propagators were
calculated on half the time extent of the MILC lattices by using a
Dirichlet boundary condition (BC) in the time direction.  With the
relatively high statistics that have now been accumulated, systematic
effects from the light states reflecting off the Dirichlet wall are
observed and are found to contaminate the correlation functions in the
region of interest (see
Fig.~\ref{fig:effMass_coarse}).  This
``lattice chopping'' strategy has been discarded, and the valence
propagators are now calculated with antiperiodic temporal BCs
imposed at the end of the full time extent of each configuration.  The
exception is on the heaviest light quark-mass point of the \coarse
ensemble.  At this heavy pion mass, the correlation function falls
sufficiently rapidly to not be significantly impacted in the region of
interest by the choice of BC.  Further, this ensemble contributes very little to our analysis in Sec.~\ref{sec:ChiPT}.

The parameters used in the present set of Lattice QCD calculations are presented in 
Table~\ref{tab:Lparams}.  
On the \coarse configurations, light quark propagators computed 
by LHPC with antiperiodic temporal BCs are used for the three lightest 
ensembles~\cite{Bratt:2010jn}.  
Strange quark propagators are computed from the same source points
in order to ``match''  the light quark propagators.
In addition, calculations on the \coarse ensembles with a lighter than
physical strange quark-mass have been performed.  
Statistics on three \fine ensembles have been accumulated, 
with the lightest pion mass being $m_\pi\approx235$~MeV.
Finally, approximately $6500$ thermalized trajectories have been completed on 
an additional rooted staggered ensemble with
the parameters
\begin{align}
&\beta=6.76,&
&bm_l^\textrm{sea}=0.007,&
&bm_s^\textrm{sea} = 0.050,&
&V = 24^3\times64\, 
\ \ ,
\end{align}
and measurements have been performed on them.
%
\begin{table}[t]
\caption{\label{tab:Lparams}
The parameters used in the Lattice QCD calculations.}
\begin{ruledtabular}
\begin{tabular}{cccccccclclc}
\multicolumn{11}{c}{$b \approx 0.125~\textrm{fm}$ ensembles}\\
$\beta$& $bm_l^\textrm{sea}$& $bm_s^\textrm{sea}$& L& T& $M_5$& $L_5$
& $bm_l^\textrm{dwf}$& $bm_l^\textrm{res}$& $bm_s^\textrm{dwf}$& $bm_s^\textrm{res}$
& $N_{src}\times N_{cfg}$\\
\hline
6.76& 0.007& 0.050& 20& 64& $1.7$& $16$&  0.0081& $0.001581(14)\footnotemark[1]$& 0.081& 0.000895(3)& $4\times 468$\\
6.76& 0.007& 0.050& 24& 64& $1.7$& $16$& 0.0081& 0.00164(3)& 0.081& 0.00091(2)& $8\times 1081$\\
6.76& 0.010& 0.050& 20& 64& $1.7$& $16$& 0.0138& $0.001566(11)\footnotemark[1]$& 0.081& 0.000913(2)& $4\times 656$\\
6.76& 0.010& 0.050& 28& 64& $1.7$& $16$& 0.0138& $0.001566(11)\footnotemark[1]$& 0.081& 0.000913(2)& $4\times 274$\\
6.79& 0.020& 0.050& 20& 64& $1.7$& $16$& 0.0313& $0.001227(11)\footnotemark[1]$& 0.081& 0.000836(3)& $4\times 486$\\
6.81& 0.030& 0.050& 20& 32& $1.7$& $16$& 0.0478& 0.001013(6)& 0.081& 0.000862(7)& $24\times 564$\\
\hline
\multicolumn{11}{c}{$b \approx 0.09~\textrm{fm}$ ensembles}\\
$\beta$& $bm_l^\textrm{sea}$& $bm_s^\textrm{sea}$& L& T& $M_5$& $L_5$
& $bm_l^\textrm{dwf}$& $bm_l^\textrm{res}$& $bm_s^\textrm{dwf}$& $bm_s^\textrm{res}$
& $N_{src}\times N_{cfg}$\\
\hline
7.08& 0.0031& 0.031& 40& 96& $1.5$& $40$& 0.0038& 0.000156(3)& 0.0423& 0.000073(2)& $1\times 170$ \\
7.08& 0.0031& 0.031& 40& 96& $1.5$& $12$& 0.0035& 0.000428(3)& 0.0423& 0.000233(2)& $1\times 422$\\
7.09& 0.0062& 0.031& 28& 96& $1.5$& $12$& 0.0080& 0.000375(4)& 0.0423& 0.000230(3)& $7\times 1001$\\
7.11& 0.0124& 0.031& 28& 96& $1.5$& $12$& 0.0164& 0.000290(3)& 0.0423& 0.000204(2)& $8\times 513$\\
\end{tabular}
\footnotetext[1]{Provided by LHPC~\cite{Bratt:2010jn}.}
\end{ruledtabular}
\end{table}

%
%
\subsection{Results of the Lattice QCD Calculations\label{sec:Latt_Results} }
\noindent
Correlation functions with the quantum numbers of the $\pi^+$ were constructed
from 
propagators generated from a gauge-invariant Gaussian-smeared source~\cite{Frommer:1995ik,Pochinsky:1997} with both smeared (SS) and point
(SP) sinks.
To determine the pion mass, the correlation functions were fit with a single
cosh toward the center of the time-direction.
\begin{align}
C_{(SX)}(t) \sim A_{(SX)}\  e^{-m_\pi T/2}\  \cosh ( m_\pi (t - T/2))\, ,
\end{align}
where $X=S,P$.
Fits incorporating excited states over larger time ranges produced consistent results for both
$m_\pi$ and $A_{(SX)}$.
With domain-wall fermions, the pion decay constant can be computed without need for operator renormalization by making use of an axial ward identity~\cite{Blum:2000kn}.
The decay constant is determined from the extracted overlap factors, $A_{(SX)}$,  along with the input quark-masses and computed values of the pion mass and residual mass, using the relation
\begin{align}
bf_\pi = \frac{A_{SP}}{\sqrt{A_{SS}}} 
\frac{2\sqrt{2} (bm_l^\textrm{dwf} + bm_l^\textrm{res})}{(bm_\pi)^{3/2}}
\, .
\end{align}
In the limit $L_5 \rightarrow \infty$, the residual chiral symmetry breaking
in the domain-wall action
vanishes 
and $m_l^\textrm{res} \rightarrow 0$.
In addition to these valence quantities, the  mixed valence-sea pion correlation functions 
have been calculated 
to extract the mixed-meson masses, as described in Ref.~\cite{Orginos:2007tw}.
%
\begin{figure}[!h]
\begin{tabular}{c}
\includegraphics[width=1.\textwidth]{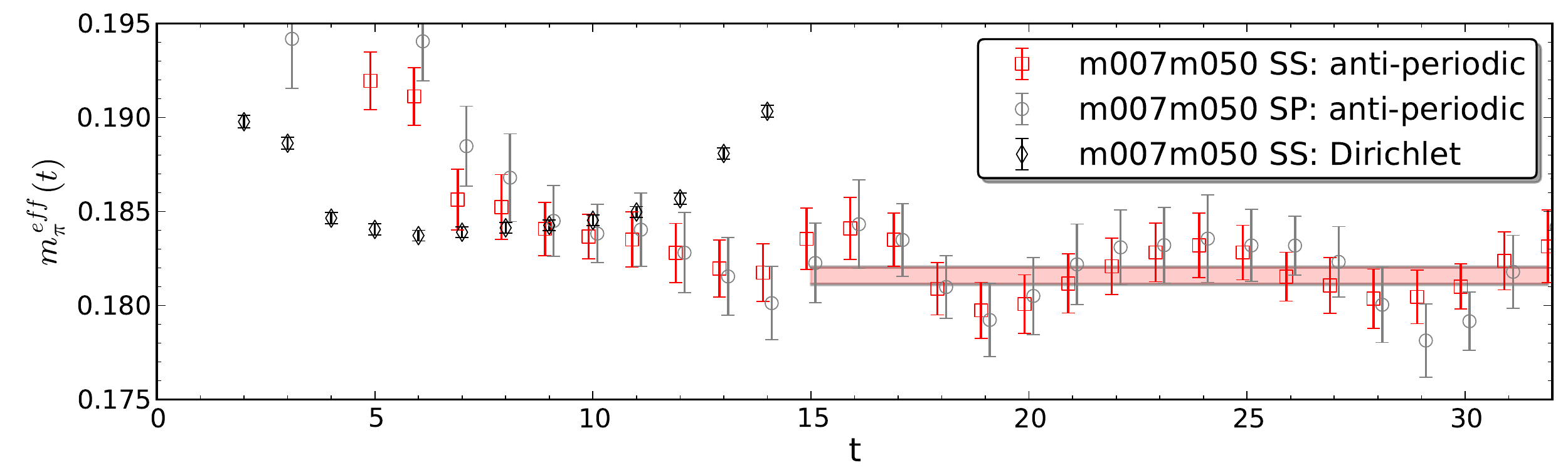}\\
\includegraphics[width=1.\textwidth]{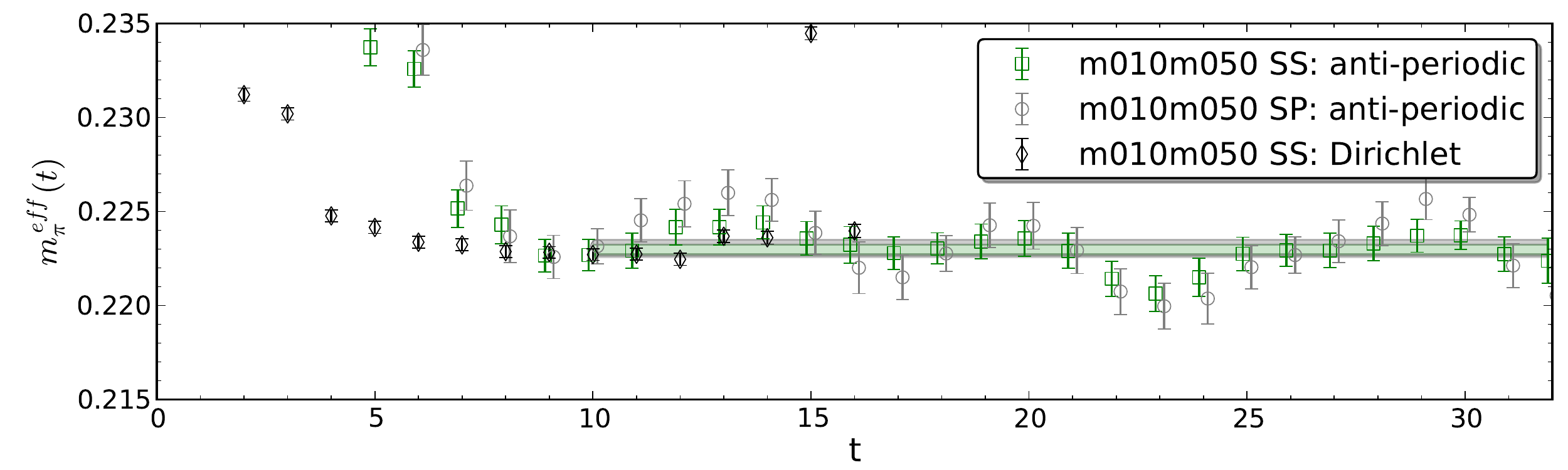}\\
\includegraphics[width=1.\textwidth]{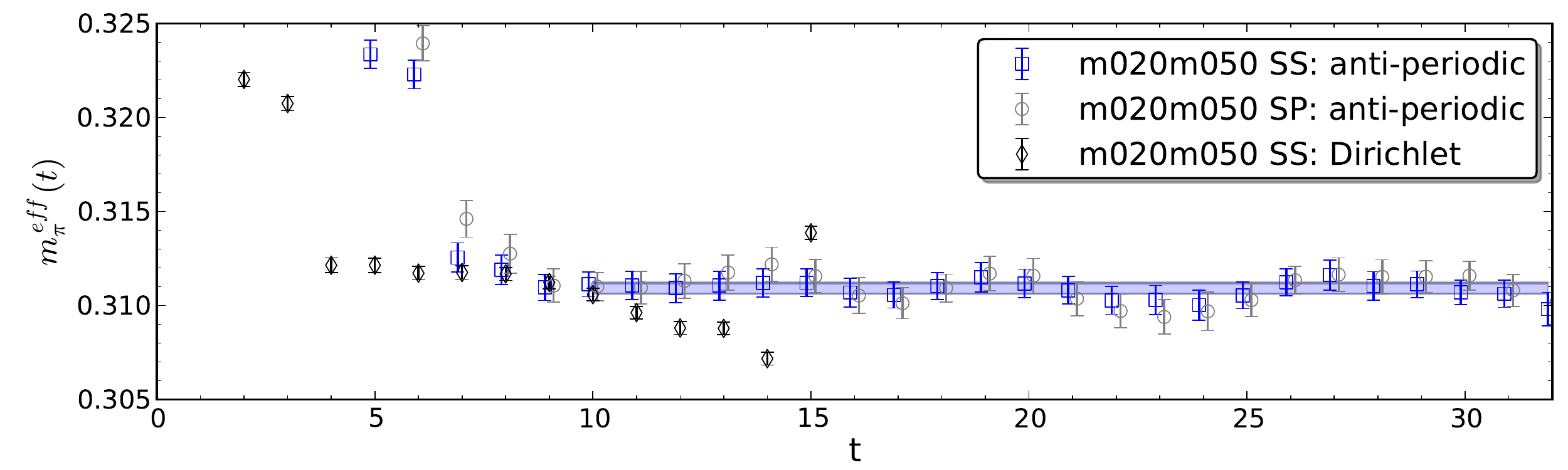}\\
\includegraphics[width=1.\textwidth]{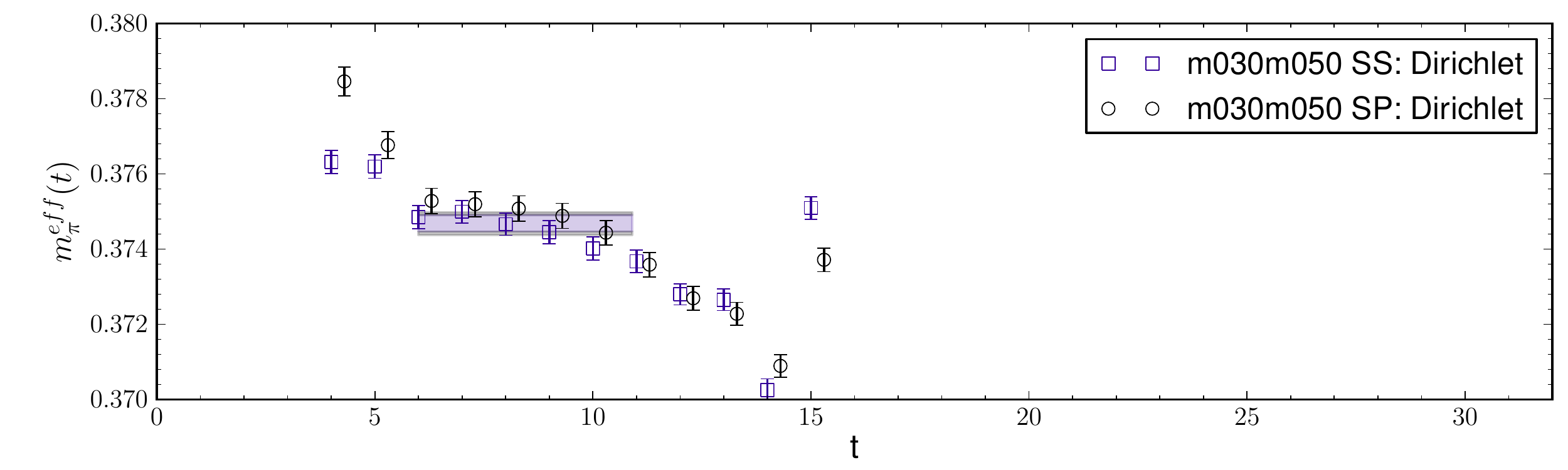}
\end{tabular}
\caption{\label{fig:effMass_coarse}EMPs of the pion correlation
    functions on 
the \coarse ensembles.  For comparative purposes, 
the effective masses from the correlation functions with Dirichlet BCs in 
time  are shown for the lightest ensembles (slightly offset for visibility).}
\end{figure}

\begin{figure}[h]
\begin{tabular}{c}
\includegraphics[width=1.\textwidth]{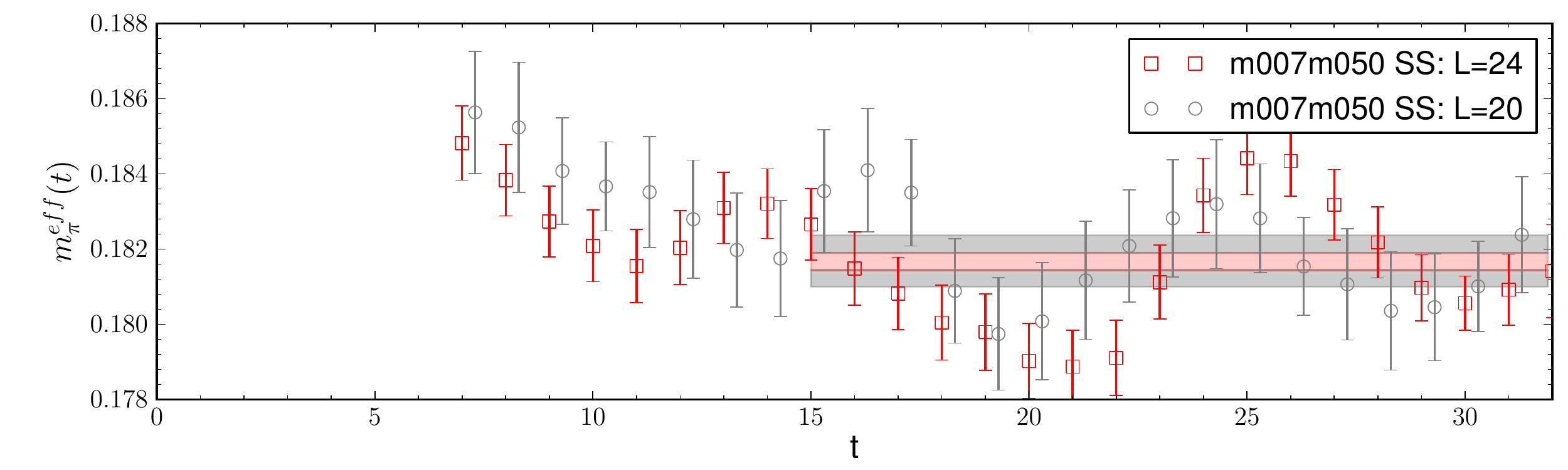}\\
\includegraphics[width=1.\textwidth]{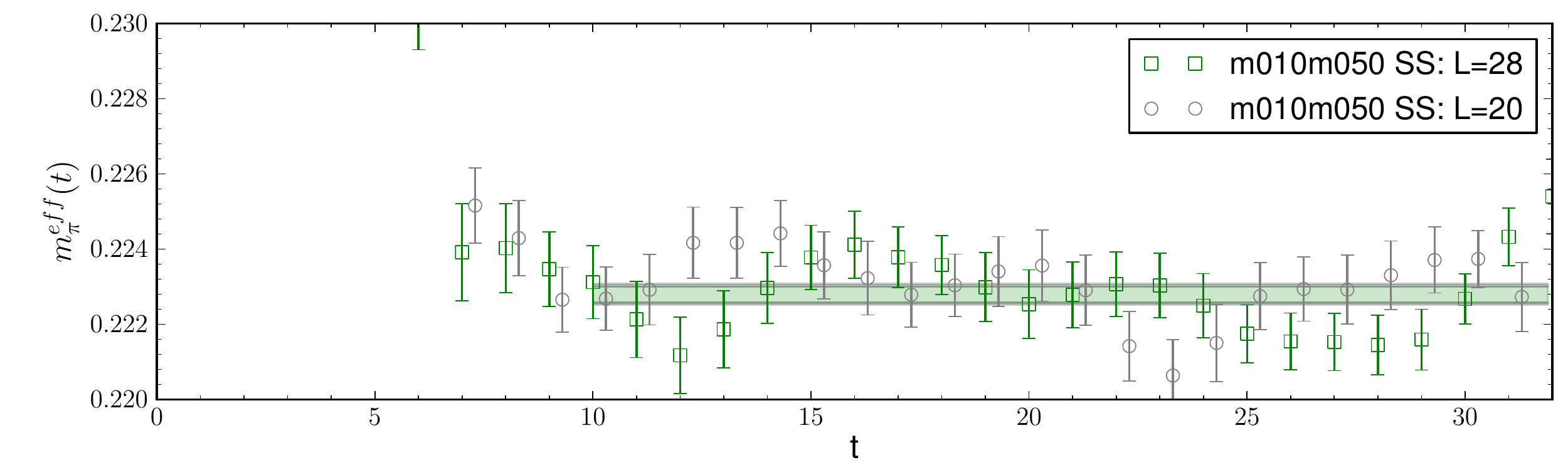}
\end{tabular}
\caption{\label{fig:effMass_coarseLV}
EMPs of the pion correlation
    functions calculated on the large volume \coarse ensembles.  
For comparative purposes,  the effective masses obtained in  the smaller volumes are
shown
(slightly offset in time for visibility).}
\end{figure}

\begin{figure}[h]
\begin{tabular}{c}
\includegraphics[width=1.\textwidth]{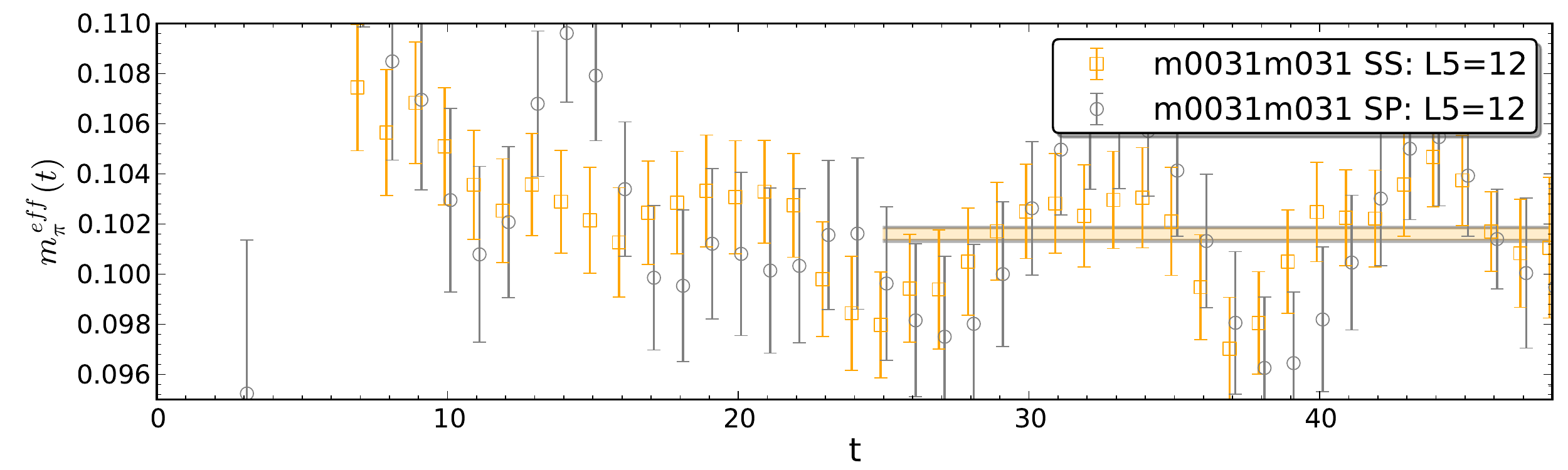}\\
\includegraphics[width=1.\textwidth]{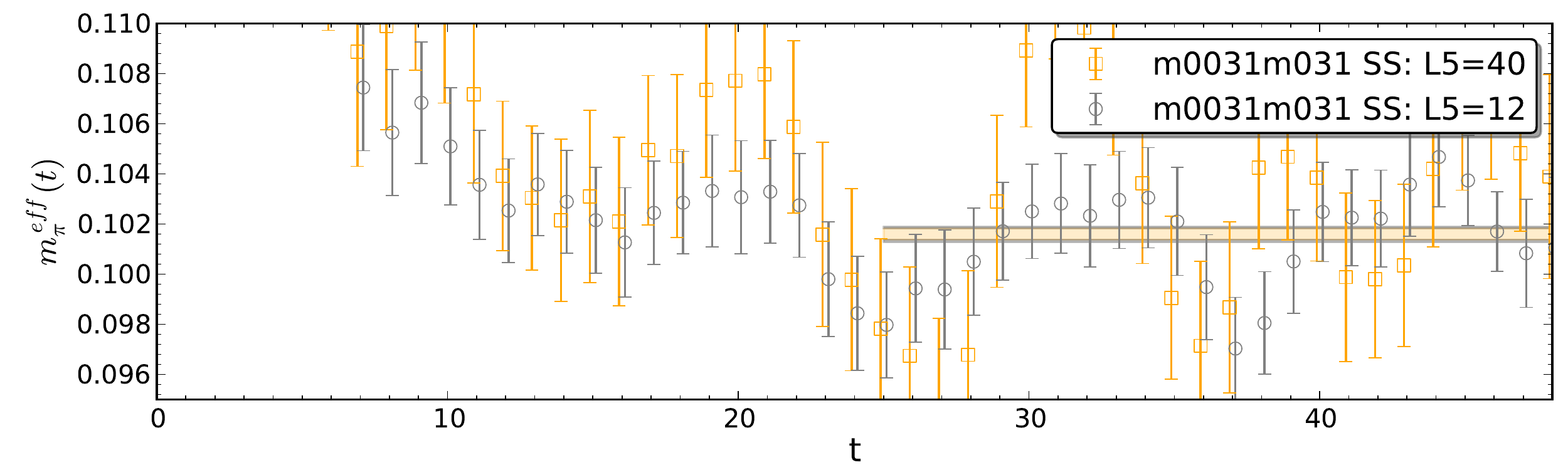}\\
\includegraphics[width=1.\textwidth]{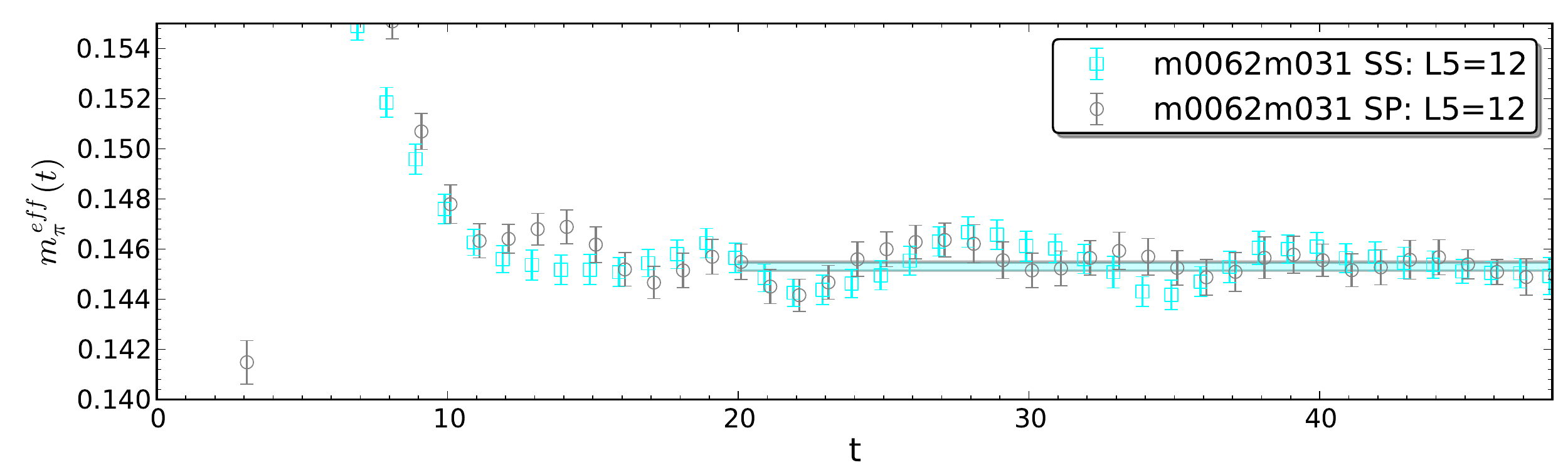}\\
\includegraphics[width=1.\textwidth]{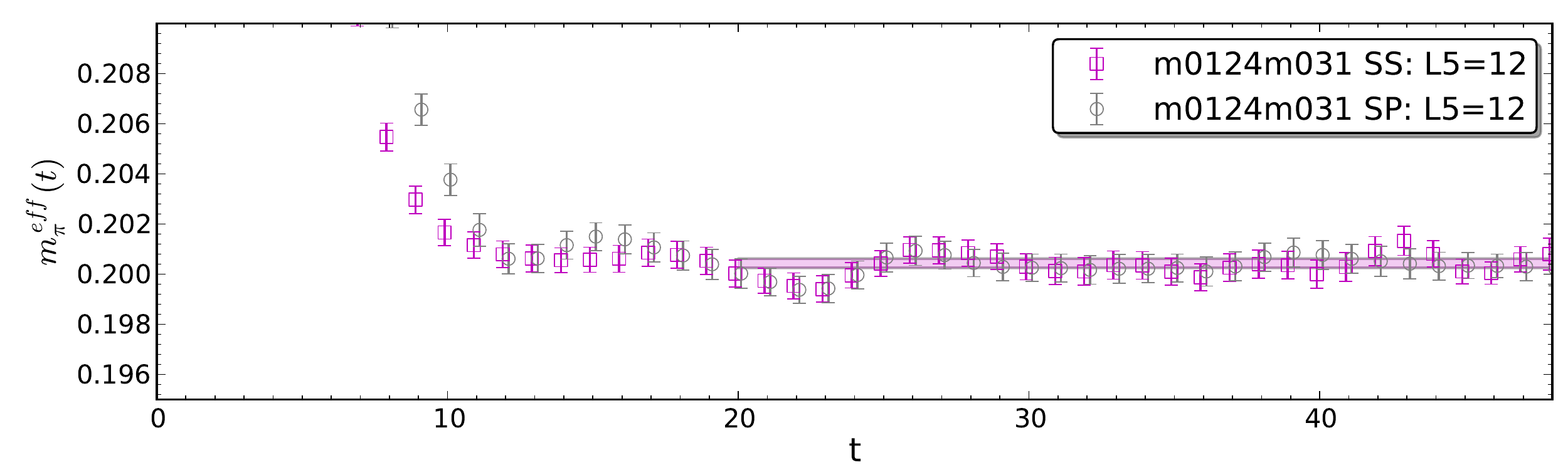}
\end{tabular}
\caption{\label{fig:effMass_fine}EMPs of the pion correlation
    functions on 
the \fine ensembles.}
\end{figure}

\begin{table}[h]
\caption{\label{tab:mpifpi_results}
The pion masses and decay constants from the Lattice QCD calculations.  
The first uncertainty is statistical and the second is systematic determined 
from the fit range.}
\begin{ruledtabular}
\begin{tabular}{cclllc}
$m_\textrm{sea}$& $L^3\times T_{val}\times L_5$& $bm_\pi$& $bf_\pi$& $bm_{\pi_{\rm Mix}}$& $m_\pi L$\\
\hline
m007m050& $20^3\times64\times16$& $0.18159(42)({}^{27}_{32})$& $0.09293(45)({}^{41}_{86})$& 0.2553(15)& 3.63 \\
m010m050& $20^3\times64\times16$& $0.22298(26)({}^{46}_{29})$& $0.09597(27)({}^{79}_{47})$& 0.2842(15)& 4.46 \\
m020m050& $20^3\times64\times16$& $0.31091(27)({}^{20}_{10})$& $0.10204(26)({}^{33}_{21})$& 0.3516(09)& 6.22 \\
m030m050& $20^3\times32\times16$& $0.37469(22)({}^{20}_{22})$& $0.10749(13)({}^{33}_{33})$& 0.412(4)& 7.49\\
\hline
m007m050& $24^3\times64\times16$& $0.18167(23)({}^{66}_{63})$ &$0.09311(28)({}^{34}_{45})$& 0.2553(15)& 4.36\\
m010m050& $28^3\times64\times16$& $0.22279(21)({}^{19}_{16})$& $0.09639(41)({}^{50}_{37})$& 0.2842(15)& 6.24\\
\hline
m0031m031& $40^3\times96\times40$& $0.10328(32)({}^{36}_{40})$& $0.0621(12)({}^{10}_{13})$& 0.1344(14)& 4.13 \\
m0031m031& $40^3\times96\times12$& $0.10160(22)({}^{21}_{24})$& $0.0617(09)({}^{10}_{13})$& 0.1293(08)& 4.06 \\
m0062m031& $28^3\times96\times12$& $0.14530(15)({}^{15}_{09})$&$0.06539(14)({}^{34}_{30})$& 0.1632(10)& 4.07\\
m0124m031& $28^3\times96\times12$& $0.20043(17)({}^{13}_{10})$& $0.07032(19)({}^{20}_{40})$& 0.2153(03)& 5.61
\end{tabular}
\end{ruledtabular}
\end{table}
%
The results of the Lattice QCD  calculations are given in Table~\ref{tab:mpifpi_results}.
Statistical uncertainties are determined from a correlated $\chi^2$ analysis as well as from a single-elimination jackknife.
Binning of the data was performed until the uncertainties did not change appreciably.
The quoted fitting systematic uncertainties are determined by varying the fit range,
including a broad sweep of $t_{min}$.
Effective mass plots (EMPs) for the
full-volume correlation functions are  generated with a \texttt{cosh}-style effective mass;
\begin{equation}\label{eq:eff_cosh}
	m_\pi^\textrm{eff} =  \frac{1}{\tau}\cosh^{-1} 
\left[ \frac{C(t + \tau) + C(t-\tau)}{2 C(t)} \right]\, ,
\end{equation}
while the others were generated with a \texttt{log}-style effective mass;
\begin{equation}
	m_\pi^\textrm{eff} = \frac{1}{\tau} \ln \left( \frac{C(t)}{C(t+\tau)} \right)\, .
\end{equation}
In Figs.~\ref{fig:effMass_coarse}-\ref{fig:effMass_fine}
the EMPs of the correlation functions and the
extracted pion masses are presented using $\tau=3$.

In Fig.~\ref{fig:effMass_coarse}, 
the effective masses from calculations with antiperiodic BCs  imposed on the valence quarks,
as well as those from the Dirichlet temporal BCs, are shown.  
Correlation functions from propagators generated with a Dirichlet BC 
(located at $t=22$ and $t=-10$ in the figures) show a significantly different
behavior from  those generated with  antiperiodic BCs.
It is for this reason that  we have abandoned the Dirichlet BC in the
generation of valence quarks.  
However, it is only the lightest ensemble on which the extracted pion 
mass determined with the Dirichlet BC 
is statistically discrepant from that generated with antiperiodic BCs.

Interestingly, the  correlation functions 
generated with antiperiodic BCs are not free of their  own systematics.  
The EMPs exhibit  an oscillation 
with a period of approximately $1$~fm, which is not simply explained by either
the staggered taste-pion mass splittings or by the mixed-meson  mass splittings.  
In the top panel of Fig.~\ref{fig:effMass_coarseLV}, the oscillations are more
pronounced 
(with higher statistics).
Comparing the EMPs from the \fine and \coarse ensembles, 
the  oscillations are seen to become more pronounced for lighter quark-masses.
As the statistics are increased, the amplitude of the oscillation becomes more significant and increasing $L_5$ does not appear to ameliorate these effects.
The choice of $\tau$ used in Eq.~\eqref{eq:eff_cosh} has no appreciable impact on the observed oscillation, unless one takes $\tau \simeq T_{osc}$, the oscillation period, in which case the oscillations are washed out.
At this point, it is not clear if the oscillations are 
an artifact of this particular mixed-action, or originate from the domain-wall
valence propagators.
Similar oscillations are observed for calculations with domain-wall valence propagators
computed on dynamical domain-wall ensembles, 
as shown in  Fig.~11 of Ref.~\cite{Aoki:2010dy} and Fig.~2 of Ref.~\cite{Aoki:2010pe}.
%
In Ref.~\cite{Aoki:1995bb}, it was suggested that these fluctuations may be explained by the time correlations in the propagators.  However, in Refs.~\cite{Beane:2009kya,Beane:2009gs,Beane:2009py,Beane:2011pc}, a calculation of the pion correlation function was performed with $\sim400$ times the number of measurements analyzed in Ref.~\cite{Aoki:1995bb}, and no evidence for such oscillations or fluctuations was found (see Figs. 17 and 18 of Ref.~\cite{Beane:2011pc}).
%
For the present work, the masses and decay constants are
determined with fits that encompass at least one full period of oscillation,
with the fitting systematic established through variations of the fitting ranges.

%
%
\subsection{Scale Setting \label{sec:Latt_Scale} }
\noindent
To extrapolate the calculated pion masses and decay
constants and make predictions at the physical pion mass, 
the scale must be determined.  
The MILC Collaboration has performed extensive scale setting analyses on their
ensembles, and  
it is used to convert the calculated pion masses and
decay constants into $r_1$ units (extrapolated to the physical values of the light quark-masses),%
\footnote{The distance $r_1$ is the Sommer scale~\cite{Sommer:1993ce} 
defined from  the heavy-quark potential at the separation,  $r_1^2 F(r_1) \equiv -1$.}
collected in Table~\ref{tab:mpifpi_r1Units}.
In Table~\ref{tab:r1b} these values are listed for the ensembles used in this
work~\cite{Bazavov:2009bb}.  
The MILC Collaboration has determined $r_1 = 0.318(7)$~fm 
using the $b\bar{b}$ meson spectrum and $r_1 = 0.311(2)({}^{3}_{8})$~fm using
$f_\pi$ to  set the scale~\cite{Bazavov:2009bb}.  
The value of 
\begin{align}\label{eq:r1_fpi}
r_1 = 0.311(2)({}^{3}_{8})\textrm{ fm}\, ,
\end{align}
is used in this work to convert to physical units.
\begin{table}
\caption{\label{tab:mpifpi_r1Units}
The pion masses (normalized to the light quark-masses) and decay constants in $r_1$ units.  The third uncertainty is the systematic from the conversion to $r_1$ units.}
\begin{ruledtabular}
\begin{tabular}{cccc}
Ensemble masses& $V$& $\frac{(r_1m_\pi)^2}{r_1 m_q}$& $r_1 f_\pi$ \\
\hline
m007m050& $20^3\times64\times16$& $\phantom{1}9.310(43)({}^{26}_{31})(11)$& $0.2545(12)({}^{11}_{23})(03)$ \\
m010m050& $20^3\times64\times16$& $\phantom{1}8.861(21)({}^{37}_{23})(10)$& $0.2628(08)({}^{23}_{14})(03)$ \\
m020m050& $20^3\times64\times16$& $\phantom{1}8.384(14)({}^{10}_{05})(10)$& $0.2879(07)({}^{09}_{06})(03)$ \\
m030m050& $20^3\times32\times16$& $\phantom{1}8.275(10)({}^{09}_{10})(12)$& $0.3093(04)(10)(05)$ \\
\hline
m007m050& $24^3\times64\times16$& $\phantom{1}9.318(23)({}^{68}_{63})(11)$& $0.2550(08)({}^{10}_{13})(03)$ \\
m010m050& $28^3\times64\times16$& $\phantom{1}8.846(16)({}^{14}_{12})(10)$& $0.2640(11)({}^{12}_{10})(03)$ \\
\hline
m0031m031& $40^3\times96\times40$& $10.123(62)({}^{70}_{78})(11)$& 0.2331(45)$({}^{38}_{49})(03)$\\
m0031m031& $40^3\times96\times12$& $\phantom{1}9.942(57)({}^{54}_{62})(11)$& 0.2318(34)$({}^{38}_{49})(03)$ \\
m0062m031& $28^3\times96\times12$& $\phantom{1}9.551(20)({}^{20}_{12})(08)$& 0.2477(05)$({}^{12}_{11})(02)$ \\
m0124m031& $28^3\times96\times12$& $\phantom{1}9.285(16)({}^{12}_{09})(10)$& 0.2713(07)$({}^{07}_{15})(03)$ \\
\end{tabular}
\end{ruledtabular}
\end{table}

\begin{table}
\caption{\label{tab:r1b}$r_1 / b$ from MILC~\cite{Bazavov:2009bb}.  
The values 
(provided by the MILC Collaboration)
extrapolated to the physical light quark-masses (rightmost
column) were used to convert from lattice units  to $r_1$ units.}
\begin{ruledtabular}
\begin{tabular}{cccc}
ensemble masses& $\beta$& $\frac{r_1}{b}(bm_l,bm_s,\beta)$& $\frac{r_1}{b}(bm_l^\textrm{phy},bm_s^\textrm{phy},\beta)$ \\
\hline
m007m050& $6.76$& 2.635(3) & 2.739(3) \\
m010m050& $6.76$& 2.618(3) & 2.739(3) \\
m020m050& $6.79$& 2.644(3) & 2.821(3) \\
m030m050& $6.81$& 2.650(4) & 2.877(4) \\
\hline
m0031m031& $7.08$& 3.695(4) & 3.755(4) \\
m0062m031& $7.09$& 3.699(3) & 3.789(3) \\
m0124m031& $7.11$& 3.712(4) & 3.858(4)
\end{tabular}
\end{ruledtabular}
\end{table}

%
%
\section{Lattice Systematics\label{sec:Sys} }
\noindent
In order to make contact with experimental measurements, the 
lattice QCD results must be extrapolated to the continuum and to infinite
volume, 
as well as to the physical 
values of the light quark-masses.  
$\chi$PT is the natural tool to perform these extrapolations, a consequence of
which is that the LECs can be determined.

%
%
\subsection{Light Quark Mass and Volume Dependence \label{sec:Sys_mq} }

\noindent
Generally, the chiral expansion at NLO involves analytic terms, chiral
logarithms and scale-dependent LECs.  
However, the perturbative expansion can be optimized 
by setting the renormalization scale to lattice-determined quantities which
vary with the quark-mass, leading to modifications at next-to-next-to-leading order (NNLO).  
For instance, the $SU(2)$ chiral expansion of $m_\pi$ and $f_\pi$ can be 
expressed as~\cite{Beane:2005rj,Colangelo:2010et}
\begin{align}
\label{eq:mpi_NLO}
m_\pi^2 &= 2 B m_q \bigg\{ 1 
	+ \frac{1}{2}\xi \ln \left( \frac{\xi}{\xi^\textrm{phy}} \right)
	-\frac{1}{2}\xi\, \bar{l}_3
	\bigg\}
\\
\label{eq:fpi_NLO}
f_\pi &= f \bigg\{ 1 
	-\xi \ln \left( \frac{\xi}{\xi^\textrm{phy}} \right)
	+\xi \bar{l}_4
	\bigg\}
\end{align}
where 
\begin{eqnarray}
\xi \ =\ {{m_\pi^2}\over 8\pi^2 f_\pi^2} \ \qquad 
{\rm and} \qquad \ \overline{l}_i = \log{\frac{\Lambda_i^2}{(m_\pi^\textrm{phy})^{2}}} \ ,
\label{eq:xidef}
\end{eqnarray}
and $\Lambda_i$ is an intrinsic scale that is not determined by chiral
symmetry.  
Here $m_\pi$ and $f_\pi$ denote lattice-measured quantities, $f$ is the
chiral-limit 
value of the pion decay constant, and $B$ is proportional to the chiral
condensate.  The ``${\rm phy}$''  superscript indicates that the relevant quantity is
evaluated with the  physical values of the pion mass and decay constant,
for which we use the
central values 
\begin{align}\label{eq:mpifpi_phys}
&f_\pi^\textrm{phy}=130.4~{\rm MeV}& 
&\textrm{and}&
&m_\pi^\textrm{phy}=139.6~{\rm MeV}\, .&
\end{align}
One benefit of performing the perturbative expansion with $\xi$ is immediately clear: as $\xi$ is dimensionless, the higher order corrections are free of scale setting ambiguities as only the LO order contributions must be expressed in terms of some lattice scale.

In addition to the light quark-mass dependence, 
the finite-volume corrections to the pion masses and decay constants
can be simply determined in the $p$-regime, defined by $m_\pi L~\gg~1$.
At NLO in the chiral expansion, 
the finite-volume corrections are given
by~\cite{Gasser:1986vb,Colangelo:2003hf}
\begin{align}
\label{eq:mpi_FV}
\DFV \frac{m_\pi^2}{2Bm_q} &= 
	8\pi^2 \D i\mc{I}(\xi,m_\pi L)
\\
\label{eq;fpi_FV}
\DFV \frac{f_\pi}{f} &= -16\pi^2 \D i\mc{I}(\xi, m_{\pi} L)
\end{align}
where
\begin{align}
8\pi^2 \D i\mc{I}(\xi,m_\pi L) &= \frac{2\xi}{m_\pi L} \sum_{n=1}^{\infty} 
\frac{k(n)}{\sqrt{n}} K_1(\sqrt{n} m_\pi L)
\end{align}
and $k(n)$ is the number of ways that the integer $n$ can be formed as the sum of squares of three integers, $n=\sum_{i=1}^3 n_i^2$ with $n_i \in \mathbb{Z}$.

The light quark-mass dependences of $m_\pi$ and $f_\pi$ are known at NNLO in 
two-flavor $\chi$PT~\cite{Bijnens:1998fm}.  
In the $\xi$ expansion, in infinite-volume, they are 
\begin{align}\label{eq:mpisq2}
\frac{m_\pi^2}{2Bm_{q}} =&\ 1 
	+\frac{1}{2}\xi \left[ \ln \left(\frac{\xi}{\xi^\textrm{phy}} \right) - \bar{l}_3 \right]
\nonumber\\&
	+\frac{7}{8} \xi^2 \ln^2 (\xi)
	-\left[ \frac{16}{3} +\frac{1}{3} \bar{l}_{12} -\frac{9}{4} \bar{l}_3 -\bar{l}_4 
		-\frac{7}{4}\ln(\xi^\textrm{phy})
	\right]
	\xi^2 \ln (\xi) 
	- \bar{l}_4\, \xi \xi^\textrm{phy}
	+\xi^2 k_M
\end{align}
and
\begin{align}\label{eq:fpi2}
\frac{f_\pi}{f} =&\ 1
	+\xi \left[ \bar{l}_4 - \ln \left(\frac{\xi}{\xi^\textrm{phy}} \right) \right]
\nonumber\\&
	+\frac{5}{4} \xi^2 \ln^2 (\xi)
	+\xi^2 \ln(\xi) \left[
		\frac{53}{12} + \frac{1}{6} \bar{l}_{12} - 5\bar{l}_4
		-\frac{5}{2} \ln (\xi^\textrm{phy})
		\right]
	+2\bar{l}_4\, \xi \xi^\textrm{phy} 
	+\xi^2 k_F
\end{align}
where $\bar{l}_{12} = 7\bar{l}_1 +8\bar{l}_2$.

%
%
\subsection{Mixed-Action $\chi$PT\label{sec:Sys_MA} }

\noindent
The low-energy EFT for mixed-action Lattice QCD calculations is well 
understood~\cite{Bar:2002nr,Bar:2003mh,Bar:2005tu,Golterman:2005xa,Tiburzi:2005is,Chen:2005ab,Prelovsek:2005rf,Aubin:2006hg,Chen:2006wf,Jiang:2007sn,Orginos:2007tw,Chen:2007ug,Aubin:2008wk,Chen:2009su}.
In Refs.~\cite{Chen:2006wf,Chen:2007ug,Chen:2009su}, 
it was demonstrated that the formulae for the pion mass and decay constant at NLO, including discretization effects, are the same for all sea-quark discretizations provided the valence quarks satisfy the Ginsparg-Wilson relation~\cite{Ginsparg:1981bj}
(including our MA approach with domain-wall valence propagators computed on rooted staggered sea-quark configurations).
The difference between the various sea-quark actions will be encoded in the values of the unphysical parameters which quantify the discretization effects.
At NLO in the MA expansion, including finite-volume effects, the pion mass and decay constant are given by
\begin{align}
\label{eq:mpi_MA_NLO}
\frac{m_\pi^2}{2Bm_q} =&\ 1 
	+ \frac{1}{2}\xi \ln \left( \frac{\xi}{\xi^\textrm{phy}} \right)
	-\frac{1}{2}\xi\, \bar{l}_3
\nonumber\\&
	-\frac{1}{2}\left(\tilde{\xi}_{\rm sea} - \xi \right) \left[ 1 + \ln \left( \xi \right) \right]
	-l_3^{PQ}\, (\xi_{\rm sea} -\xi)
	+l_3^{b}\, \left( \frac{b}{r_1} \right)^2
\nonumber\\&
	+8\pi^2 \D i\mc{I}(\xi,m_\pi L)
	+8\pi^2 (\tilde{\xi}_{\rm sea} - \xi )\D \partial i\mc{I}(m_\pi L)\, ,
\\ \nonumber\\
\label{eq:fpi_MA_NLO}
\frac{f_\pi}{f} =&\ 1 
	-\tilde{\xi}_{\rm Mix} \ln \left( \frac{\tilde{\xi}_{\rm Mix}}{\xi^\textrm{phy}} \right)
	+\xi \bar{l}_4
\nonumber\\&
	-\left( \tilde{\xi}_{\rm Mix} - \xi \right) \ln \left( \xi^\textrm{phy} \right)
	-l_4^{PQ}\, (\xi_{\rm sea} -\xi)
	+l_4^{b}\, \left( \frac{b}{r_1} \right)^2
\nonumber\\&
	-16\pi^2 \D i\mc{I}(\tilde{\xi}_{\rm Mix}, m_{\pi_{\rm Mix}} L)\, ,
\end{align}
where
\begin{align}
\D\partial i\mc{I}(m L) = \frac{1}{(4\pi)^2} \sum_{n=1}^{\infty} 
	k(n) \left(
			K_0(\sqrt{n}mL) 
			+K_2(\sqrt{n} mL)
			- \frac{2 K_1(\sqrt{n}mL)}{\sqrt{n}mL} 
		\right)
\end{align}
\begin{table}
\caption{\label{tab:xi_mix}Expansion parameters $m_l / m_s$, $\xi$, $\tilde{\xi}_{\rm Mix}$, $\tilde{\xi}_{\rm sea} - \xi$, $\xi_{\rm sea} -\xi$ and $\frac{m^\textrm{res}}{m_q}$.}
\begin{ruledtabular}
\begin{tabular}{cccccccc}
$m_\textrm{sea}$& V& $m_l / m_s$& $\xi$& $\tilde{\xi}_{\rm Mix}$ & $\tilde{\xi}_{\rm sea} - \xi$& $\xi_{\rm sea} - \xi$& $\frac{m^\textrm{res}}{m_q}$ \\
\hline
m007m050& $20^3\times64\times16$& 0.14& $0.0491$& $0.096$& $0.114$& 0.0032& 0.165 \\
m010m050& $20^3\times64\times16$& 0.20& $0.0681$& $0.111$& $0.108$& 0.0010& 0.102 \\
m020m050& $20^3\times64\times16$& 0.40& $0.1177$& $0.150$& $0.093$& 0.0001& 0.038 \\
m030m050& $20^3\times32\times16$& 0.60& $0.1540$& $0.186$& $0.084$& 0.0026& 0.021 \\
\hline
m007m050& $24^3\times64\times16$& 0.14& $0.0489$& $0.096$& $0.114$& 0.0032& 0.165 \\
m010m050& $28^3\times64\times16$& 0.20& $0.0674$& $0.111$& $0.108$& 0.0010& 0.102 \\
\hline
m0031m031& $40^3\times96\times40$& 0.10& $0.0360$& 0.058& 0.050& 0.0004& 0.039\\
m0031m031& $40^3\times96\times12$& 0.10& $0.0365$& $0.058$& $0.050$& 0.0004& 0.109 \\
m0062m031& $28^3\times96\times12$& 0.20& $0.0629$& $0.079$& $0.045$& 0.0019& 0.045 \\
m0124m031& $28^3\times96\times12$& 0.40& $0.1037$& $0.119$& $0.038$& 0.0054& 0.017
\end{tabular}
\end{ruledtabular}
\end{table}
For the present calculations, the extra expansion parameters of the theory are defined as
\begin{align}
&\tilde{\xi}_{\rm Mix} = 
	\frac{\frac{1}{2}\left(m_{\pi}^2 + m_{\pi_{\rm sea,5}}^2\right) + b^2 \Delta^\prime_{\rm Mix}}{8\pi^2 f_\pi^2}&
&\tilde{\xi}_{\rm sea} = \frac{m_{\pi_{\rm sea,5}}^2 + b^2 \Delta_{\rm I}}{8\pi^2 f_\pi^2}&
&\xi_{\rm sea} = \frac{m_{\pi_{\rm sea,5}}^2}{8\pi^2 f_\pi^2}&
\end{align}
where $m_{\pi_{\rm sea,5}}$ is the taste-5 staggered pion mass, $b^2 \D_{\rm
  I}$ is 
the mass splitting of the taste identity staggered pion and $b^2 \D^\prime_{\rm
  Mix}$ 
is the mass splitting of the mixed valence-sea
pion~\cite{Bar:2005tu,Chen:2009su}, 
determined in Refs.~\cite{Orginos:2007tw,Aubin:2008wk} and this work.
In Table~\ref{tab:xi_mix}, the values of the parameters relevant for the 
calculations are listed.

\begin{figure}
\begin{tabular}{cc}
\includegraphics[width=0.48\textwidth]{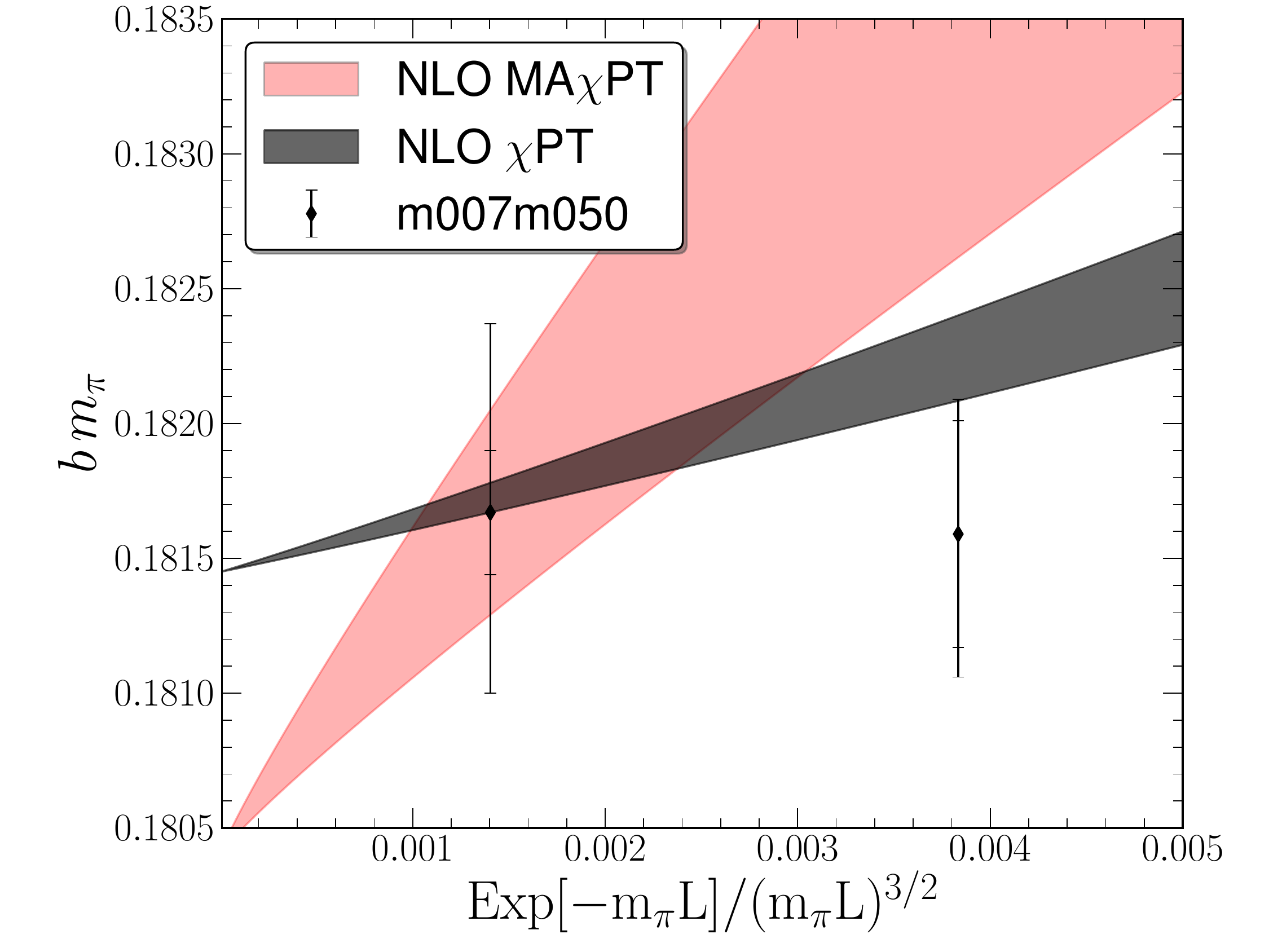}
&
\includegraphics[width=0.48\textwidth]{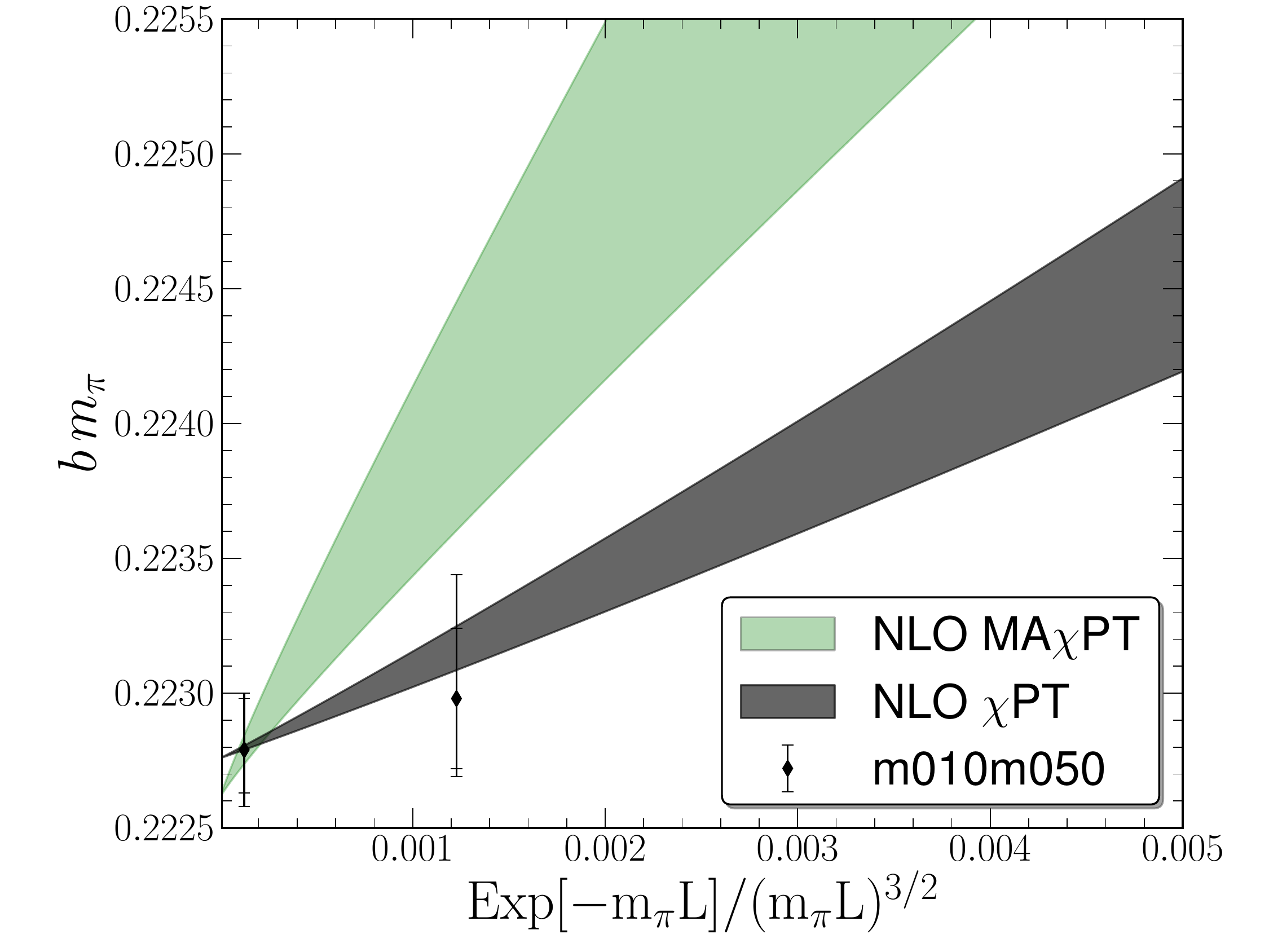}
\\
\includegraphics[width=0.48\textwidth]{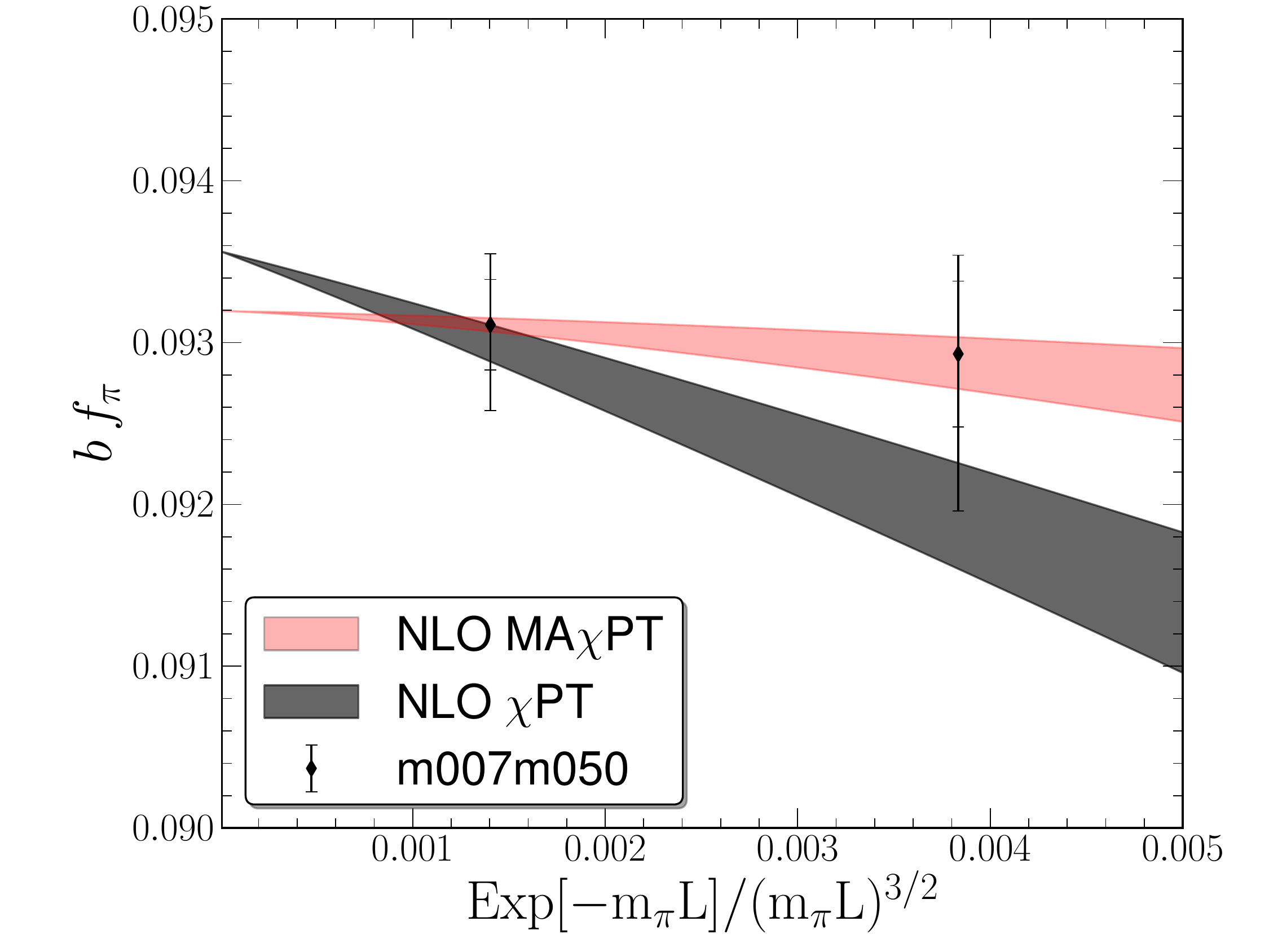}
&
\includegraphics[width=0.48\textwidth]{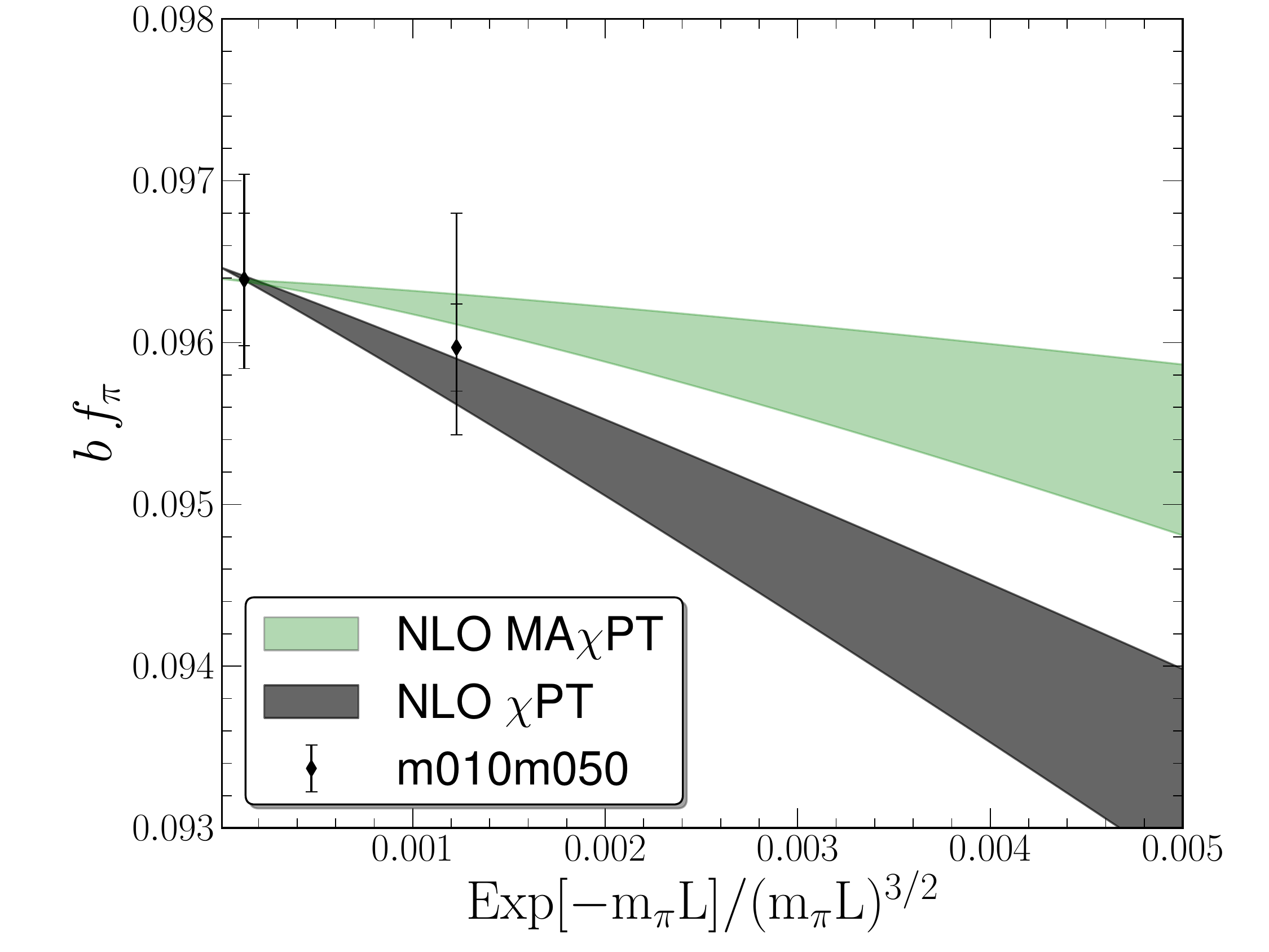}
\end{tabular}
\caption{\label{fig:mpiFV}NLO finite-volume contributions, and an estimate of their uncertainty, 
in $\chi$PT and MA$\chi$PT compared with the results of the Lattice QCD
calculations on the m007m050 and m010m050 ensembles.  
The central values have been chosen to coincide for the larger volume ensembles.
}
\end{figure}

In analogy with finite-volume $\chi$PT,  
the pion mass and pion decay constant in finite-volume MA$\chi$PT are 
related to their infinite-volume values at NLO via the relations
\begin{align}
\label{eq:mpiFV}
m_\pi[FV] = m_\pi \bigg\{ 1
	&+\frac{1}{2} \sum_{n=1}^\infty \frac{k(n)}{2} \bigg[
		4\xi \frac{K_1(\sqrt{n} m_\pi L)}{\sqrt{n} m_\pi L}
\nonumber\\&
		+(\xi_\textrm{sea} - \xi) \left(
			K_0(\sqrt{n} m_\pi L) + K_2(\sqrt{n} m_\pi L) - 2 \frac{K_1(\sqrt{n} m_\pi L)}{\sqrt{n} m_\pi L}
		\right)
	\bigg]
	\bigg\}\, ,
\end{align}
and
\begin{align}
\label{eq:fpiFV}
f_\pi[FV] = f_\pi \left[
		1
		-4 \xi_\textrm{Mix} \sum_{n=1}^\infty k(n) 
\frac{K_1(\sqrt{n} m_{\pi_\textrm{Mix}} L)}{\sqrt{n} m_{\pi_\textrm{Mix}} L}
	\right]\, .
\end{align}
In the case of $f_\pi$, 
the finite-volume effects in MA$\chi$PT are somewhat suppressed compared to
those in $\chi$PT.  This is because the contribution from the ``average''
valence-sea type virtual pion in a one-loop diagram is smaller than from a 
valence-valence pion due to its larger mass~\cite{Orginos:2007tw}.
In contrast, the pion mass receives a one-loop contribution from a hairpin
diagram~\cite{Sharpe:1997by}, which has enhanced volume effects compared to a
typical one-loop contribution.  
In Table~\ref{tab:mpifpiFV}, the FV contributions to 
$m_\pi$ and $f_\pi$ from  Eqs.~\eqref{eq:mpiFV} and Eq.~\eqref{eq:fpiFV} are presented.
%
\begin{table}
\caption{\label{tab:mpifpiFV}
Finite-volume corrections to $m_\pi$ and $f_\pi$ at NLO in MA$\chi$PT, as given
in Eqs.~\eqref{eq:mpiFV} and Eq.~\eqref{eq:fpiFV}.
For a quantity $Y$ in the table, $\delta Y[FV]/Y = (Y[FV]-Y)/Y$.
}
\begin{ruledtabular}
\begin{tabular}{c|cccccc}
& \multicolumn{6}{c}{\coarse ensemble} \\
Quantity & \multicolumn{2}{c}{m007m050}& \multicolumn{2}{c}{m010m050}& m020m050& m030m050\\
& $L=20$& $L=24$& $L=20$& $L=28$& $L=20$& $L=20$\\
\hline
MA$\chi$PT: $\delta m_\pi[FV] / m_\pi$& 1.6\%& 0.6\%& 0.6\%& 0.1\%& 0.1\%& 0.0\% \\
$\chi$PT: $\delta m_\pi[FV] / m_\pi$& 0.2\%& 0.1\%& 0.1\%& 0.0\%& 0.0\%& 0.0\% \\
MA$\chi$PT: $\delta f_\pi[FV] / f_\pi$& -0.3\%& -0.1\%& -0.2\%& -0.0\%& -0.1\%& -0.0\% \\
$\chi$PT: $\delta f_\pi[FV] / f_\pi$& -1.4\%& -0.5\%& -0.6\%& -0.1\%& -0.1\%& -0.0\% \\
\hline\hline
& \multicolumn{6}{c}{\fine ensemble } \\
Quantity & m0031m031& & m0062m031& & m0124m031\\
& $L=40$& & $L=28$& & $L=28$ \\
\hline
MA$\chi$PT: $\delta m_\pi[FV] / m_\pi$& 0.4\%& & 0.4\%&&  0.1\% \\
$\chi$PT: $\delta m_\pi[FV] / m_\pi$& 0.1\%&&  0.1\%&&  0.0\% \\
MA$\chi$PT: $\delta f_\pi[FV] / f_\pi$& -0.2\%&&  -0.6\%&&  -0.1\% \\
$\chi$PT: $\delta f_\pi[FV] / f_\pi$& -0.6\%&&  -0.9\%&&  -0.2\%
\end{tabular}
\end{ruledtabular}
\end{table}
%
On the lightest two coarse ensembles, the NLO volume contributions  to $m_\pi$
from MA$\chi$PT   are substantially
larger than those from  $\chi$PT.
Further, due to the high precision of the Lattice QCD calculations, the
finite-volume volume contributions are larger than the
uncertainties on the m007m050 ensembles.
This is in contrast to the results of the Lattice QCD calculations of  $m_\pi$, 
which show little volume dependence.
In Ref.~\cite{Colangelo:2005gd}, it was demonstrated that NNLO $\chi$PT
could increase  the finite-volume contributions by as much as $\sim 50\%$ of
the NLO contribution.
In the case of MA$\chi$PT, with hairpin diagrams having enhanced volume
effects, the importance of the NNLO contributions is likely to 
be even greater than in $\chi$PT.
As these NNLO effects have not yet been calculated, the MA$\chi$PT
finite-volume contributions are assigned a $30\%$ systematic uncertainty when
performing the analysis in  Sec.~\ref{sec:ChiPT}.  
In Fig.~\ref{fig:mpiFV},  the NLO finite-volume contributions in $\chi$PT and
in MA$\chi$PT for the m007m050 and m010m050 ensembles are compared with the
results of the Lattice QCD calculations.
The $\chi$PT band is given by the range $\D m_\pi = (1+0.5)\D
m_\pi^\textrm{$\chi$PT}$,  while the MA$\chi$PT corrections are given by 
$\D m_\pi = (1 \pm 0.3) \D m_\pi^\textrm{MA$\chi$PT}$,
where the central values have been chosen to coincide for the larger volume ensembles.
The MA$\chi$PT finite-volume contributions appear not to describe the observed
volume dependence of $m_\pi$,  
indicating the likely importance of NNLO contributions.
In the case of $f_\pi$, the volume contributions are in good agreement with
the results of the Lattice QCD calculations.

%
%
\subsection{Strange Quark Mass Effects \label{sec:Sys_ms} }
\noindent
The strange quark-masses used in the present  calculations are not
equal to the  physical 
value~\cite{Aubin:2004ck}; the physical staggered strange quark-mass was
determined 
to be $bm_s^\textrm{phy} = 0.0350(7)$ and $bm_s^\textrm{phy} = 0.0261(5)$ on 
the $b\approx0.125$~\textrm{fm} and $b\approx 0.09$~\textrm{fm} ensembles 
respectively~\cite{Bazavov:2009bb}.  
In order to estimate the effects of this small mistuning in the two-flavor expansion, a matching to $SU(3)$ $\chi$PT must be performed, where it is found the effects can be absorbed into the NLO LECs ~\cite{Gasser:1984gg};
%
\begin{align}
\label{eq:lbar_ms_corrections}
&\bar{l}_3(m_s,m_s^\textrm{phy}) = \bar{l}_3(m_s^\textrm{phy}) + \d \bar{l}_3(m_s,m_s^\textrm{phy})\, ,&
&\d \bar{l}_3(m_s,m_s^\textrm{phy}) = -\frac{1}{9} \ln \left( \frac{m_s}{m_s^\textrm{phy}} \right)&
\nonumber\\
&\bar{l}_4(m_s,m_s^\textrm{phy}) = \bar{l}_4(m_s^\textrm{phy}) + \d \bar{l}_4(m_s,m_s^\textrm{phy})\, ,&
&\d \bar{l}_4(m_s,m_s^\textrm{phy}) = \frac{1}{4} \ln \left( \frac{m_s}{m_s^\textrm{phy}} \right)&
\end{align}
These lead to mild corrections to $\bar{l}_3$ and $\bar{l}_4$ on both the coarse and fine ensembles,
\begin{align}
\d \bar{l}_3(m_s,m_s^\textrm{phy}) &= \left\{\begin{array}{llr}
		-0.040(2),& b\approx 0.125 \textrm{ fm}, bm_s^\textrm{sea}=0.05\\
		-0.019(1),& b\approx 0.09\phantom{5} \textrm{ fm}, bm_s^\textrm{sea}=0.031 \ ;\\
	\end{array}\right. 
\nonumber\\ \nonumber\\
\d \bar{l}_4(m_s,m_s^\textrm{phy}) &= \left\{\begin{array}{ll}
		\phantom{+}0.089(5),& b\approx 0.125 \textrm{ fm}, bm_s^\textrm{sea}=0.05\\
		\phantom{+}0.043(5),& b\approx 0.09\phantom{5} \textrm{ fm},
                bm_s^\textrm{sea}=0.031
\ \ .
\\
	\end{array}\right.
\end{align}
These strange quark-mass mistuning effects are negligible compared with the uncertainties of the extracted values for $\bar{l}_3$ and $\bar{l}_4$ (see Sec.~\ref{sec:ChiPT}).

%
%
\subsection{Residual Chiral Symmetry Breaking Effects \label{sec:Sys_mres} }

\noindent 
The domain-wall action has residual chiral symmetry breaking due to
the finite extent of the fifth dimension, $L_5$, 
resulting from the overlap of the chiral modes bound to opposite walls in the fifth-dimension.
The  quantity $m^\textrm{res}$ is the leading manifestation of this residual
chiral symmetry breaking, and the effective quark-mass of the Lattice QCD
calculation becomes 
\begin{equation}\label{eq:mq_mres}
m_q = m_{l}^\textrm{dwf} + m_l^\textrm{res}\, ,
\end{equation}
capturing the dominant effects of the residual chiral symmetry breaking
appearing at LO in the chiral Lagrangian.  
However, it is known that there are subleading effects.  
Defining the quark-mass through Eq.~\eqref{eq:mq_mres} and taking the 
standard definition of $m^\textrm{res}$ as the ratio of two pion to vacuum matrix elements~\cite{Blum:2000kn}
\begin{equation}
bm^\textrm{res} \equiv \frac{\langle 0 | J^a_{5q} | \pi \rangle}{\langle 0| J^a_5 | \pi \rangle}\, ,
\end{equation}
where $J^a_{5q}$ and $J^a_5$ are pseudoscalar densities made, respectively, from quarks in the middle and boundaries of the fifth dimension,
the quantity $m^\textrm{res} = m^\textrm{res}(bm_{l},b)$ depends upon the input
quark-mass and the  lattice-spacing 
(see Ref.~\cite{Aoki:2010dy} for a discussion of these effects).  
Consequently, the chiral Lagrangian receives a simple modification at 
NLO~\cite{Blum:2001xb,Golterman:2005fe,Sharpe:2007yd}.
Following the method of Ref.~\cite{Sharpe:1998xm}, 
the modifications to the chiral Lagrangian at NLO are
\begin{align}
\d \mc{L}_\textrm{res} & = \ 
	\frac{l_3^\textrm{res} + l_4^\textrm{res}}{16} \tr \left( 2B m_q \S +
          2B m_q \S^\dagger \right) 
\tr \left( 2B m^\textrm{res} \S + 2B m^\textrm{res} \S^\dagger \right)
\nonumber\\ 
	& +\ 
\frac{l_4^\textrm{res}}{8} \tr \left( \partial_\mu \S \partial^\mu \S^\dagger
\right) 
\tr \left( 2B m^\textrm{res} \S + 2B m^\textrm{res} \S^\dagger \right)\, .
\label{eq:newops}
\end{align}
The corrections to $m_\pi$ and $f_\pi$ arising from these new terms are 
\begin{eqnarray}
\label{eq:mpifpi_mres_corrections}
\frac{\d m_\pi^2}{2Bm_q} & =&  -\frac{1}{2}\xi\,   \frac{m^\textrm{res}}{m_q}\, \bar{l}_3^\textrm{res}
\qquad {\rm and }\qquad
\frac{\d f_\pi}{f}\ =\  \xi\, \frac{m^\textrm{res}}{m_q}\, 
\bar{l}_4^\textrm{res}
\ \ \ ,
\end{eqnarray}
with
\begin{align}
\bar{l}_i^\textrm{res} = \frac{32\pi^2}{\g_i}\, l_i^\textrm{res}\, ,
\end{align}
where $\g_3 = -1/2$ and $\g_4 = 2$~\cite{Gasser:1983yg}.
As with the coefficients $l_i^b$, these $l_i^\textrm{res}$ coefficients are not universal and depend upon the choice of lattice action used.

The new operators in Eq.~(\ref{eq:newops})
were found to give the
dominant uncertainty in the prediction of the $I=2\ \pi\pi$ scattering length
at the physical pion mass~\cite{Beane:2007xs} as the $l_i^\textrm{res}$ were unknown.  
Therefore, for $\pi\pi$ scattering, and for other observables, it is important
to determine the $l_i^\textrm{res}$, which can be done simply by
performing calculations with different values of $L_5$ on the same ensemble .  
The fine MILC ensembles, with $b\approx0.09$~fm, at the lightest quark-mass
point were used to perform calculations with $L_5 = 12$ and $L_5 =
40$.
\begin{table}
\caption{\label{tab:mres}
Parameters used to isolate  $m^\textrm{res}$ effects.  
The $L_5 = 16,24$ calculations were used to tune the quark-mass for the  $L_5 = 40$ calculation
in such a way that the sum $b(m_{l} + m_l^\textrm{res})$ was the same (within $\sim
0.7\%$) for the $L_5=12$
and $40$ calculations.}
\begin{ruledtabular}
\begin{tabular}{ccccccc}
Ensemble& $L_5$& $b m_{l}$& $b m^\textrm{res}$& $\frac{m^\textrm{res}}{m_{l} + m^\textrm{res}}$&$bm_\pi$& $bf_\pi$\\
\hline
4096f21b708m0031m031
& 12& 0.0035& 0.000428(03)& 0.109(1)& $0.10160(22)({}^{21}_{24})$& $0.0617(12)({}^{10}_{13})$\\
& 16 & 0.0030& 0.000321(11)& 0.0987(3) & - & -  \\
& 24 & 0.0030& 0.000229(12)& 0.071(4) &  - & - \\
& 40& 0.0038& 0.000156(03)& 0.039(1)& $0.10328(32)({}^{36}_{40})$& $0.0621(09)({}^{10}_{13})$\\
\end{tabular}
\end{ruledtabular}
\end{table}
The quark-mass, defined by Eq.~\eqref{eq:mq_mres}, was tuned to be the same for
both $L_5$'s, which was achieved to within 0.7\% accuracy (giving the same
value of $m_\pi^2$ up to $\sim 3\%$).
The results of the calculations are presented in Table~\ref{tab:mres}.
The values of $l_3^\textrm{res}$ and $l_4^\textrm{res}$ that are determined by
the Lattice QCD calculations are presented in Sec.~\ref{sec:ChiPT}.

%
%
\section{Chiral, Continuum and Volume Extrapolations \label{sec:ChiPT} }

\noindent
The numerical results presented in this work were obtained at several values of the light quark-masses and two lattice spacings.  To control the discretization effects, it would be ideal to have at least three lattice spacings: however, a third smaller lattice-spacing is beyond the scope of this work.
To address this limitation, the chiral and continuum extrapolations are performed in two different ways.
%
The first method is to fit the LECs of $\chi$PT to the \coarse and \fine
calculations  independently.  
The extracted LECs are then extrapolated to the continuum limit, using the ansatz%
\footnote{The leading discretization corrections in the current formulation of MA lattice QCD scale as $\mc{O}(b^2)$.}
\begin{equation}\label{eq:LEC_cont}
\lambda(b) = \lambda_0 + \lambda_2 \left(\frac{b}{r_1}\right)^2 \, .
\end{equation}
This analysis is performed at both NLO and NNLO in the chiral expansion.  
The second method  to perform the continuum and chiral extrapolations is to use
MA$\chi$PT, which leads to determinations of the LECs that are
consistent with those obtained with the first method.
This lends confidence that the  discretization effects are small enough to be
captured  by the MA$\chi$PT formulation.

Before proceeding, it should be noted that the 
light quark-masses are given in lattice units and have not been converted to a
continuum regularization scheme.  
As the product $m_q B$ is renormalization scheme and scale independent, the
values of the LEC B, which we determine, have not been properly converted to a continuum regularization scheme.  For this reason, we do not provide the results of this quantity.

%
%
\subsection{Method 1: $\chi$PT  and Continuum Extrapolation
\label{sec:ChiPT_Vanilla} }

%
%
\subsubsection{NLO $SU(2)$ \label{sec:ChiPT_Vanilla_NLO} }
\noindent
The pion masses and decay constants obtained in 
the Lattice QCD calculations  on the \coarse and \fine ensembles are
used to determine the LECs at NLO in $\chi$PT by independently 
fitting to the expressions in  Eqs.~\eqref{eq:mpi_NLO} and \eqref{eq:fpi_NLO}, 
including the FV corrections in  Eqs.~\eqref{eq:mpi_FV} and \eqref{eq;fpi_FV}.
Strange quark-mass effects are included by using
Eq.~\eqref{eq:lbar_ms_corrections}, but 
residual chiral symmetry breaking effects, such as those described by
Eq.~\eqref{eq:mpifpi_mres_corrections}, are not.
Both the mass and decay constant depend upon two LECs each, as seen from 
Eqs.~\eqref{eq:mpi_NLO} and \eqref{eq:fpi_NLO}.
The uncertainties in the values of $\xi$ and other parameters in Table~\ref{tab:xi_mix} are included in our analysis through our Monte Carlo treatment but do not appreciably impact the analysis.
Including the larger volume calculations, the complete set of results presented in 
Table~\ref{tab:mpifpi_r1Units} utilizes six data sets on the \coarse ensembles and three on the \fine ensembles.
For each of the NLO fixed lattice-spacing fits that are presented in
Tables~\ref{tab:mpi_NLO_Vanilla} and \ref{tab:fpi_NLO_Vanilla},
the maximum value of $m_l / m_s$ used in the fit is listed.
On the \coarse ensembles, the ratio is in the range  $m_l / m_s = 0.14 - 0.6$, while
on the  \fine ensembles the ratio is in the range $m_l / m_s = 0.1 -
0.4$%
\footnote{
In addition to giving the $\chi^2$ and the number of 
degrees of freedom ($dof$) in the fit, 
the $Q$-value, or \textit{confidence of fit}, is also provided,
\begin{equation}\label{eq:Q}
Q \equiv \int_{\chi^2_{min}}^{\infty} d \chi^2\ \mc{P}(\chi^2,d)\, ,
\end{equation}
where 
\begin{equation}
	\mc{P}(\chi^2,d) = \frac{1}{2^{d/2} \G(d/2)} (\chi^2)^{d/2 - 1} e^{-\chi^2 / 2}
\end{equation}
is the probability distribution function for $\chi^2$ with $d$ degrees of freedom.
(The $Q$-value represents the probability that if a random sampling of data were
taken from the parent distribution, a larger $\chi^2$ would result.)
}.
%
\begin{table}
\caption{\label{tab:mpi_NLO_Vanilla}
Results of the  fixed lattice-spacing NLO $\chi$PT analysis of $m_\pi$.
Max $m_l/m_s$ denotes the maximum value of the ratio of light quark-masses used
to perform the analysis.
}
\begin{ruledtabular}
\begin{tabular}{c|cccc}
Max& \multicolumn{4}{c}{\coarse}\\ 
$m_l/m_s$& $\bar{l}_3$& $\chi^2_{stat+sys}$& $dof$& $Q$ \\
\hline
0.4& 5.09(06)(52)& 18.1& 3& 0.00 \\
0.6& 4.60(03)(36)& 46.6& 4& 0.00 \\
\hline\hline
& \multicolumn{4}{c}{\fine}\\ 
$m_l/m_s$& $\bar{l}_3$& $\chi^2_{stat+sys}$& $dof$& $Q$ \\
\hline
0.4& $4.05(10)(40)$& 3.31& 1& 0.07 \\
\end{tabular}
\end{ruledtabular}
\end{table}
%
%
\begin{table}
\caption{\label{tab:fpi_NLO_Vanilla}
Results of the fixed lattice-spacing NLO $\chi$PT analysis of $f_\pi$.
Max $m_l/m_s$ denotes the maximum value of the ratio of light quark-masses used
to perform the analysis.
}
\begin{ruledtabular}
\begin{tabular}{c|ccccc}
Max&\multicolumn{5}{c}{\coarse}\\ 
$m_l/m_s$& $r_1 f$& $\bar{l}_4$& $\chi^2_{stat+sys}$& $dof$& $Q$ \\
\hline
0.4& 0.2166(10)$(40)$& 4.78(06)$(20)$& 2.35& 3& 0.50 \\
0.6& 0.2109(07)$(13)$& 5.28(03)$(10)$& 15.3& 4& 0.00\\
\hline\hline
&\multicolumn{5}{c}{\fine}\\
$m_l/m_s$ & $r_1 f$& $\bar{l}_4$& $\chi^2_{stat+sys}$& $dof$& $Q$\\
\hline
0.4& 0.1983(16)(34)& 5.48(13)(28)& 0.15& 1& 0.69
\end{tabular}
\end{ruledtabular}
\end{table}
%

From the quality of fit given in Tables~\ref{tab:mpi_NLO_Vanilla} and ~\ref{tab:fpi_NLO_Vanilla}, 
it is clear that the NLO $\chi$PT formula for $m_\pi$ fails to
describe the results of the Lattice QCD calculation at either lattice-spacing, while the NLO $\chi$PT formula for $f_\pi$ describes the 
results on the lightest three \coarse ensembles  well and describes all the
results on the \fine ensembles.  
Taking the results of the fits with $m_l / m_s \leq 0.4$,
a continuum extrapolation of the extracted LECs using Eq.~\eqref{eq:LEC_cont}
gives
\begin{align}
&\bar{l}_3 = 3.2(0.2)(1.2)&
&\textrm{and}&
&\bar{l}_4 = 6.3(0.3)(1.1)\, .&
\end{align}
The NLO $\chi$PT 
determination of $\bar{l}_3$ must be taken with extreme caution (and essentially discarded)
as the fit to $m_\pi$ is poor.
This (relatively) large value of $\bar{l}_4$ extracted at NLO is consistent with the JLQCD NLO results using
$n_f=2$ overlap fermions~\cite{Noaki:2008iy}.

%
%
\subsubsection{NNLO $SU(2)$ \label{sec:ChiPT_Vanilla_NNLO} }

\noindent
The pion mass and decay constant at NNLO in $\chi$PT, given in 
Eq.~\eqref{eq:mpisq2} and Eq.~\eqref{eq:fpi2}, 
depend upon two additional LECs, $k_M$ and $k_F$, in addition to the appearance
of further NLO LECs $\bar{l}_{12} = 7\bar{l}_1 + 8\bar{l}_2$.
Both $\bar{l}_1$ and $\bar{l}_2$ are reasonably well determined from $\pi\pi$ 
scattering~\cite{Colangelo:2001df},
\begin{align}
\label{eq:l1l2}
&\bar{l}_1 =  -0.4(6)&
&\textrm{and}&
&\bar{l}_2 \ = \ 4.3(1)\, .&
\end{align}
To perform the fits at NNLO, these values of $\bar{l}_{1}$ and $\bar{l}_2$
are used as input.
Normal distributions of $\bar{l}_1$ and $\bar{l}_2$ are generated with means and variances given by Eq.~\eqref{eq:l1l2}, 
which are then used in the fitting process. This allows for a determination of 
the systematic uncertainty generated by their use as input parameters.
In fitting to the results of the calculations on the \fine ensembles,
there are six Lattice QCD results, and six fit parameters.
The results of this analysis are collected in Table~\ref{tab:NNLO_Vanilla}.
\begin{table}
\caption{\label{tab:NNLO_Vanilla}
Results of the  continuum NNLO $\chi$PT analysis of $m_\pi$ and $f_\pi$.
}
\begin{ruledtabular}
\begin{tabular}{c|ccccc|ccc}
Max&\multicolumn{8}{c}{\coarse}\\ 
$m_l/m_s$& $r_1 f$& $\bar{l}_3$& $\bar{l}_4$& $k_M$& $k_F$& $\chi^2_{stat+sys}$& $dof$& $Q$ \\
\hline
0.4& 0.233(04)(08)& 7.95(35)(60)& 2.63(37)(67)& 29(3)(4)& 21(6)(10)& 0.53& 4& 0.74 \\
0.6& 0.230(02)(03)& 5.83(14)(18)& 2.95(14)(24)& 14(1)(1)& 16(2)(3)& 10.0& 6& 0.12 \\
\hline
&\multicolumn{8}{c}{\fine}\\
& $r_1 f$& $\bar{l}_3$& $\bar{l}_4$& $k_M$& $k_F$& $\chi^2_{stat+sys}$& $dof$& $Q$ \\
\hline
0.4& 0.203(11)(15)& 5.61(67)(73)& 4.1(1.1)(1.6)& 19(5)(5)& 2(17)(25)& 0& 0& --\\
\end{tabular}
\end{ruledtabular}
\end{table}
%
The NNLO $\chi$PT is found to describe the results of the Lattice QCD
calculations for both $m_\pi$ and $f_\pi$.
Taking  the \coarse and \fine fit and using them to perform a continuum extrapolation,
\begin{align}
&\bar{l}_3 = 3.3(1.4)(1.7)&
&\textrm{and}&
&\bar{l}_4 = 5.8(2.4)(3.5)&
\end{align}
are obtained, consistent with those from the NLO analysis.  
These results must also be treated with caution 
due to the  small number of calculations performed on the \fine ensembles.
In Figs.~\ref{fig:l3} and \ref{fig:l4}, one can see the approximate contribution of discretization effects in the values of $\bar{l}_3$ and $\bar{l}_4$.

%
%
\subsection{Method 2: Mixed-Action $\chi$PT  \label{sec:ChiPT_MA} }

\noindent
As in the continuum case, the $m_\pi$ and $f_\pi$ analyses with  MA$\chi$PT are
decoupled at NLO in the expansion, but 
the results of the Lattice QCD calculations at both lattice spacings
can be fit simultaneously.  
This allows for several choices of fit ranges, which are denoted as  A-E in Table~\ref{tab:AtoE}.  
The maximum value of $m_l / m_s$ used in the fits from the \coarse and \fine
ensembles are listed in Table~\ref{tab:AtoE}.
As discussed in Sec.~\ref{sec:Sys_MA}, 
the NLO MA$\chi$PT volume contributions are assigned a $30\%$ uncertainty as an
estimate of NNLO effects.  This additional uncertainty is combined 
in quadrature with the other quoted systematic uncertainties.
\begin{table}
\caption{\label{tab:AtoE} 
Fit ranges used in the MA$\chi$PT analysis.  
For a given fit, A--E, the maximum value of $m_l / m_s$ (sea-quark-masses)
is given.}
\begin{tabular}{cccc}
\hline\hline
Fit & \multicolumn{3}{c}{Max $m_l / m_s$} \\
\hline
& COARSE &  COARSE & FINE \\
& $L=20$ & $L=24,28$ & \\
\hline
A& 0.20& 0.20& 0.20 \\
B& 0.20& 0.20& 0.40 \\
C& 0.40& 0.20& 0.20 \\
D& 0.40& 0.20& 0.40 \\
E& 0.60& 0.20& 0.40 \\
\hline\hline
\end{tabular}
\end{table}

%
%
\subsubsection{NLO Mixed-Action $\chi$PT \label{sec:ChiPT_MA_NLO} }

\begin{table}
\caption{\label{tab:NLO_MA_mpi}Results from NLO MA$\chi$PT fits to $(r_1 m_\pi)^2 / (r_1 m_q)$.}
\begin{ruledtabular}
\begin{tabular}{c|cccc|ccc}
&\multicolumn{6}{c}{LECs}\\
Fit& $\bar{l}_3$& $l_3^{b}$& $\bar{l}_3^\textrm{res}$& $l_3^{PQ}$& $\chi^2_{stat+sys}$& $dof$& $Q$ \\
\hline
A& $4.27(23)({}^{36}_{39})$& $-1.23(21)({}^{25}_{29})$& $14(6)({}^{7}_{8})$& $-0.6(1.6)({}^{2.8}_{2.3})$& 1.41& 2& 0.49 \\
B& $4.11(21)({}^{29}_{38})$& $-1.09(19)({}^{20}_{34})$& $19(5)({}^{5}_{9})$& $-2.9(0.9)({}^{2.0}_{1.4})$& 2.33& 3& 0.51 \\
C& $4.10(19)({}^{21}_{27})$& $-1.16(20)({}^{20}_{34})$& $17(6)({}^{5}_{9})$& $-1.4(1.5)({}^{3.5}_{1.7})$& 1.78& 3& 0.62 \\
D& $4.10(19)({}^{21}_{28})$& $-1.09(19)({}^{19}_{34})$& $19(5)({}^{5}_{9})$& $-2.8(0.8)({}^{1.4}_{0.8})$& 2.33& 4& 0.67 \\
E& $4.10(19)({}^{21}_{28})$& $-1.13(18)({}^{18}_{30})$& $18(5)({}^{5}_{8})$& $-2.7(0.7)({}^{1.1}_{0.7})$& 2.36& 5& 0.80 \\
\end{tabular}
\end{ruledtabular}
\end{table}

\begin{table}
\caption{\label{tab:NLO_MA_fpi}Results from NLO MA$\chi$PT fits to $r_1 f_\pi$.}
\begin{ruledtabular}
\begin{tabular}{c|ccccc|ccc}
&\multicolumn{6}{c}{LECs}\\
Fit& $r_1 f$& $\bar{l}_4$& $l_4^{b}$& $\bar{l}_4^\textrm{res}$& $l_4^{PQ}$& $\chi^2_{stat+sys}$& $dof$& $Q$ \\
\hline
A& $0.1847(61)({}^{80}_{89})$& $5.80(52)({}^{68}_{54})$& $0.6(0.9)({}^{1.0}_{1.1})$& $-2(12)({}^{15}_{13})$& $-3.8(5.5)({}^{8.7}_{7.3})$& 0.27& 2& 0.87 \\
B& $0.1860(20)({}^{36}_{51})$& $5.73(42)({}^{55}_{39})$& $0.5(0.8)({}^{0.8}_{0.9})$& $-1(11)({}^{12}_{11})$& $-2.7(2.6)({}^{4.4}_{3.2})$& 0.28& 3& 0.96 \\
C& $0.1812(26)({}^{55}_{36})$& $6.03(40)({}^{38}_{43})$& $0.8(0.8)({}^{0.8}_{1.0})$& $-5(12)({}^{14}_{11})$& $-6.1(4.4)({}^{8.3}_{5.0})$& 0.32& 3& 0.96 \\
D& $0.1841(17)({}^{33}_{39})$& $5.99(39)({}^{39}_{41})$& $0.4(0.8)({}^{0.9}_{0.8})$& $\phantom{-}1(11)({}^{11}_{12})$& $-0.9(2.4)({}^{3.3}_{3.7})$& 0.58& 4& 0.97 \\
E& $0.1797(12)({}^{24}_{31})$& $6.10(40)({}^{36}_{45})$& $0.9(0.8)({}^{0.9}_{0.8})$& $-5(11)({}^{11}_{12})$& $-2.9(2.4)({}^{2.4}_{4.2})$& 3.48& 5& 0.63 \\
\end{tabular}
\end{ruledtabular}
\end{table}

\noindent
Fits are performed over the ranges listed in Table~\ref{tab:AtoE},
the results of these analyses are collected in Table~\ref{tab:NLO_MA_mpi} and 
Table~\ref{tab:NLO_MA_fpi}.
There are a few observations to make.  
First, the NLO MA$\chi$PT formula is capable of describing the results of the
Lattice QCD calculations of  $m_\pi$, unlike the NLO $\chi$PT formula.  
Second, the MA$\chi$PT provides a slightly better description of the pion decay
constant than of the pion mass.  
In both cases, the NLO formula is capable of describing the results of the Lattice
QCD calculations over  the full range of quark-masses.

As the $Q$-value has a probabilistic interpretation, it is convenient to use it
in forming weighted averages of the quantities that have been extracted with
multiple fitting procedures and/or different numbers of degrees of freedom.
For extractions of a parameter $\lambda$ from different
procedures, each giving $\lambda_i$ with $Q_i$, the weighted average
\begin{equation}
\bar{\lambda} = \frac{\sum_i Q_i \lambda_i}{\sum_j Q_j}\, ,
\end{equation}
can be formed.%
\footnote{
NPLQCD has consistently performed systematic uncertainty analysis by weighting the results of different but equivalent fitting strategies~\cite{Beane:2005rj,Beane:2006mx,Beane:2006pt,Beane:2006fk,Beane:2006kx,Beane:2006gj,Beane:2006gf,Beane:2007xs,Beane:2007uh,Beane:2007es,Detmold:2008fn,Beane:2008dv,Detmold:2008yn,Detmold:2008bw,Torok:2009dg}.   
This particular method of $Q$-weighting has also been advocated by the BMW Collaboration~\cite{Durr:2008zz}, for example.
}
As each of the fits considered in this work, presented in Table~\ref{tab:AtoE},
includes successively larger  quark-masses, 
this averaging will give more weight to the lighter quark-mass
values,  where $\chi$PT is more reliable.
Performing this $Q$-weighted averaging of the results from Tables~\ref{tab:NLO_MA_mpi} and \ref{tab:NLO_MA_fpi} gives
\begin{align}
\label{eq:l3l4_NLOMA}
	&\bar{l}_3[NLO] =  4.13(20)({}^{25}_{31})\, ,&
	&\bar{l}_4[NLO] = 6.09(40)({}^{37}_{45})\, ,&
\nonumber\\
	&\bar{l}_3^\textrm{res}[NLO] = 18(5)({}^{5}_{9})\, ,&
	&\bar{l}_4^\textrm{res}[NLO] = -5(11)({}^{11}_{12})\, .&
\end{align}
The value of $\bar{l}_3$ is consistent with the average of all other Lattice
QCD calculations~\cite{Colangelo:2010et}.  
However, the value of $\bar{l}_4$ is noticeably higher, but 
is consistent with that obtained with $N_f=2$ overlap
fermions and 
a NLO $\chi$PT analysis~\cite{Noaki:2008iy}.
While the residual chiral symmetry
breaking LECs are not well determined, they will help constrain the
analysis of the $I=2\ \pi\pi$ scattering length~\cite{Beane:2007xs}.

%
%
\subsubsection{NLO MA$\chi$PT  + NNLO $SU(2)$ $\chi$PT 
\label{sec:ChiPT_MA_NNLO} }
\noindent
While the complete NNLO expressions for the pion mass and decay constant are
not available in MA$\chi$PT, it is useful to consider the  hybrid construction of
NLO MA$\chi$PT plus NNLO $\chi$PT.
As in the previous section, the NLO MA$\chi$PT volume contributions are
assigned a 30\% uncertainty.
Further, the infinite-volume formulae for the NNLO contributions are used.
While the fit values of the NNLO LECs 
will be polluted by discretization effects, 
the NLO Gasser-Leutwyler coefficients will be 
be free of these contaminations, and further, their extracted values 
should be stabilized with the inclusion of these higher order contributions.

\begin{table}
\caption{\label{tab:NNLO_MA}
Extracted values of the LECs from NLO MA$\chi$PT plus NNLO $\chi$PT fitting of the Lattice QCD results.
Data set A has insufficient light quark-mass range to constrain the NNLO analysis.
}
\begin{ruledtabular}
\begin{tabular}{c|ccccc|ccc}
&\multicolumn{5}{c}{LECs}\\
Fit& $r_1 f$& $\bar{l}_3$& $\bar{l}_4$& $k_M$& $k_F$& $\chi^2_{stat+sys}$& $dof$& $Q$ \\
\hline
A& --& --& --& --& --& --& --& -- \\
B& $0.186(9)(13)$& $4.48(51)({}^{89}_{77})$& $4.83(94)({}^{1.4}_{1.3})$ 
	&$13(5)({}^{8}_{7})$& $-8(17)({}^{25}_{24})$& 2.22& 4& 0.69 \\
C& $0.188(7)({}^{\ 9}_{11})$& $4.12(30)({}^{57}_{71})$& $4.38(55)({}^{89}_{65})$
	&$\phantom{1}8(2)({}^{4}_{5})$& $1(8)({}^{10}_{13})$& 2.17& 4& 0.70 \\
D& $0.193(5)({}^{\ 5}_{10})$& $4.00(28)({}^{77}_{53})$& $4.10(44)({}^{87}_{45})$
	&$\phantom{1}6(2)({}^{6}_{3})$& $5(6)({}^{\ 7}_{13})$& 2.99& 6& 0.81 \\
E& $0.194(3)({}^{\ 5}_{\ 7})$& $3.69(14)({}^{18}_{19})$& $4.01(22)({}^{36}_{24})$
	& $\phantom{1}3(1)(1)$& $7(2)({}^{\ 3}_{\ 4})$& 3.63& 8& 0.89 \\
\end{tabular}
\end{ruledtabular}
\end{table}
The fit functions for $m_\pi$ and $f_\pi$ share two LECs;
at NNLO, $m_\pi^2$ depends upon $\bar{l}_4$ as well as $\bar{l}_3$, and both depend upon $\bar{l}_{12}$, see Eqs.~\eqref{eq:mpisq2} and \eqref{eq:fpi2}.
In principle, a correlated analysis should be performed; however, 
the correlations only exist at NNLO, and are expected to be insignificant.
To capture the effects of the correlations on the central value of $\bar{l}_4$, the extrapolation analysis is performed with a Monte Carlo.  Further, as seen in Fig.~\ref{fig:mpifpi_expansion}, the NNLO contributions to $m_\pi$ are insignificant, supporting the above expectation.
In order to verify these expectations, a fully correlated fit was performed on a subset of the fits, A--E.  The change in the values of the LECs was well contained within the quoted uncertainties.
Results of these fits are presented in Table~\ref{tab:NNLO_MA} for the various data sets.
Taking the Q-weighted average of these results gives
\begin{align}
&\bar{l}_3[NNLO] = \lthree\, ,&
&\bar{l}_4[NNLO] = \lfour\, ,&
\nonumber\\
&\bar{l}_3^\textrm{res}[NNLO] = \lthreeRes\, ,&
&\bar{l}_4^\textrm{res}[NNLO] = \lfourRes\, .&
\end{align}
with 
$\bar{l}_3[NNLO] $ and $\bar{l}_4[NNLO] $
in good agreement with the averages given in Ref.~\cite{Colangelo:2010et}.
At NNLO in the chiral expansion, corrections to the pion decay constant are found to be 
\begin{equation}
\frac{f_\pi}{f}[NNLO] = \fpif\, .
\end{equation}
Setting the scale either by using $r_1^\textrm{phy} = 0.311(2)({}^{3}_{8})$~fm
from the MILC Collaboration to determine $f_\pi^\textrm{phy}$, 
or by using the experimental value of $f_{\pi^+}$ to determine $r_1$, gives
\begin{align}
&f_\pi^\textrm{phy}[\textrm{NNLO}] = \fpi \textrm{ MeV}&
&\textrm{and}&
&r_1^\textrm{phy}[\textrm{NNLO}] = \rone \textrm{ fm}\, .&
\end{align}
where the last uncertainty in the postdicted value of $f_\pi$ comes from MILC's
determination of $r_1$, Eq.~\eqref{eq:r1_fpi}.
%
\begin{figure}
\begin{tabular}{cc}
\includegraphics[width=0.48\textwidth]{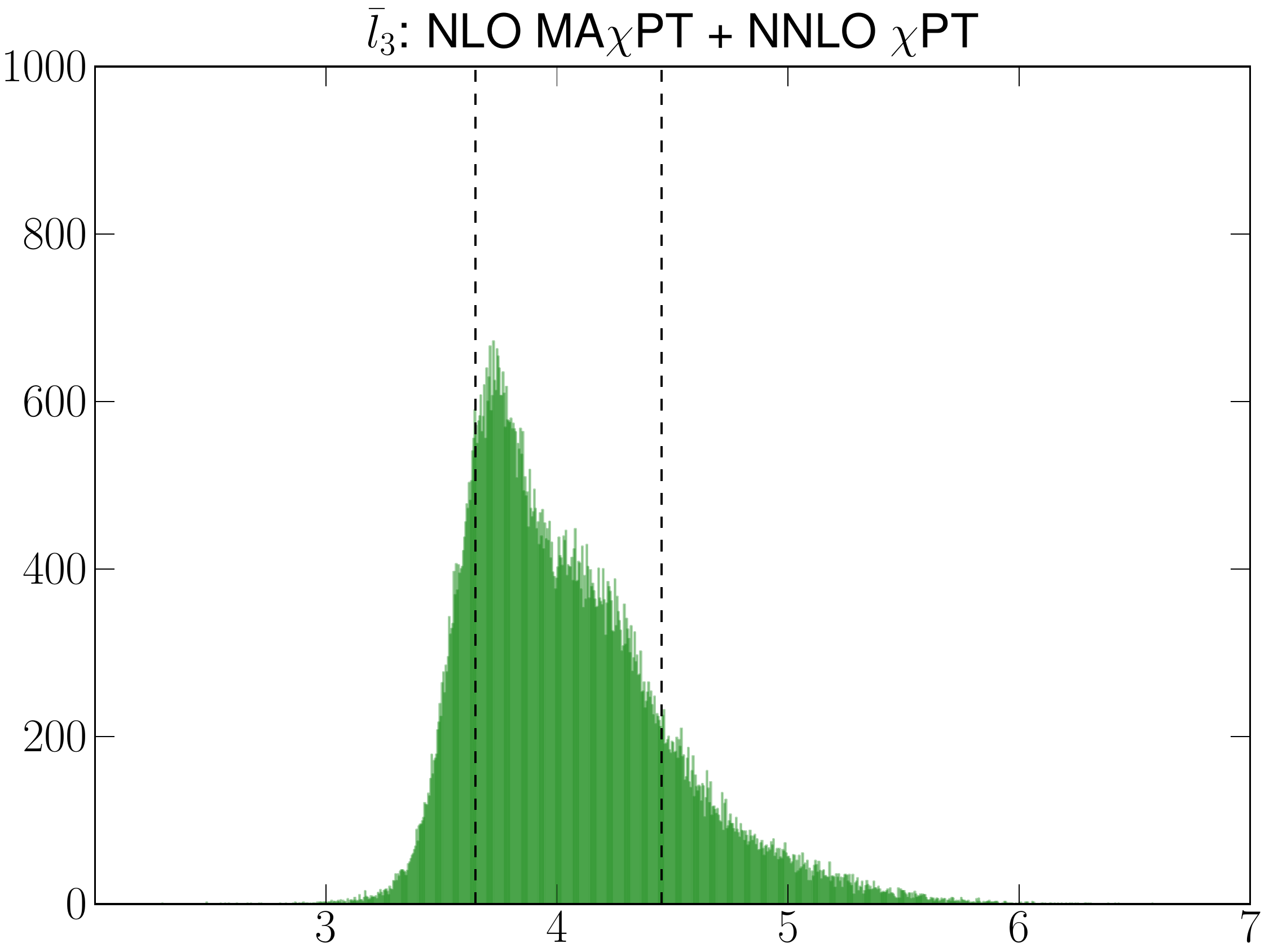}
&
\includegraphics[width=0.48\textwidth]{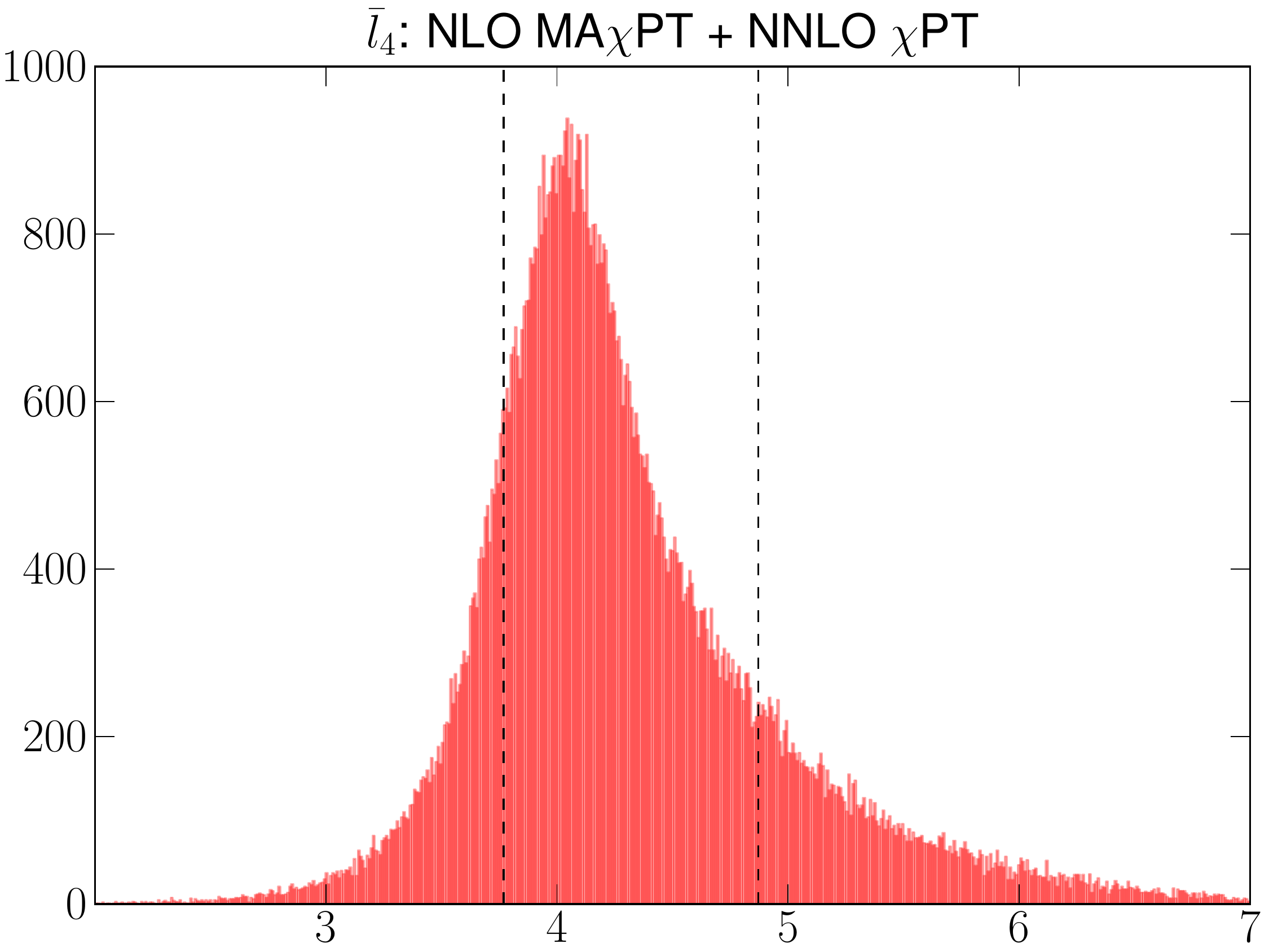}
\end{tabular}
\caption{\label{fig:l3l4hist}
$\bar{l}_3$ and $\bar{l}_4$
generated through a Monte Carlo averaging of the fits in
Table~\ref{tab:NNLO_MA}. 
The histograms are generated with $10^5$ samplings.  The vertical dashed lines represent the 16\% and 84\% quantiles.
}
\end{figure}
%

\begin{figure}
\begin{tabular}{cc}
\includegraphics[width=0.75\textwidth]{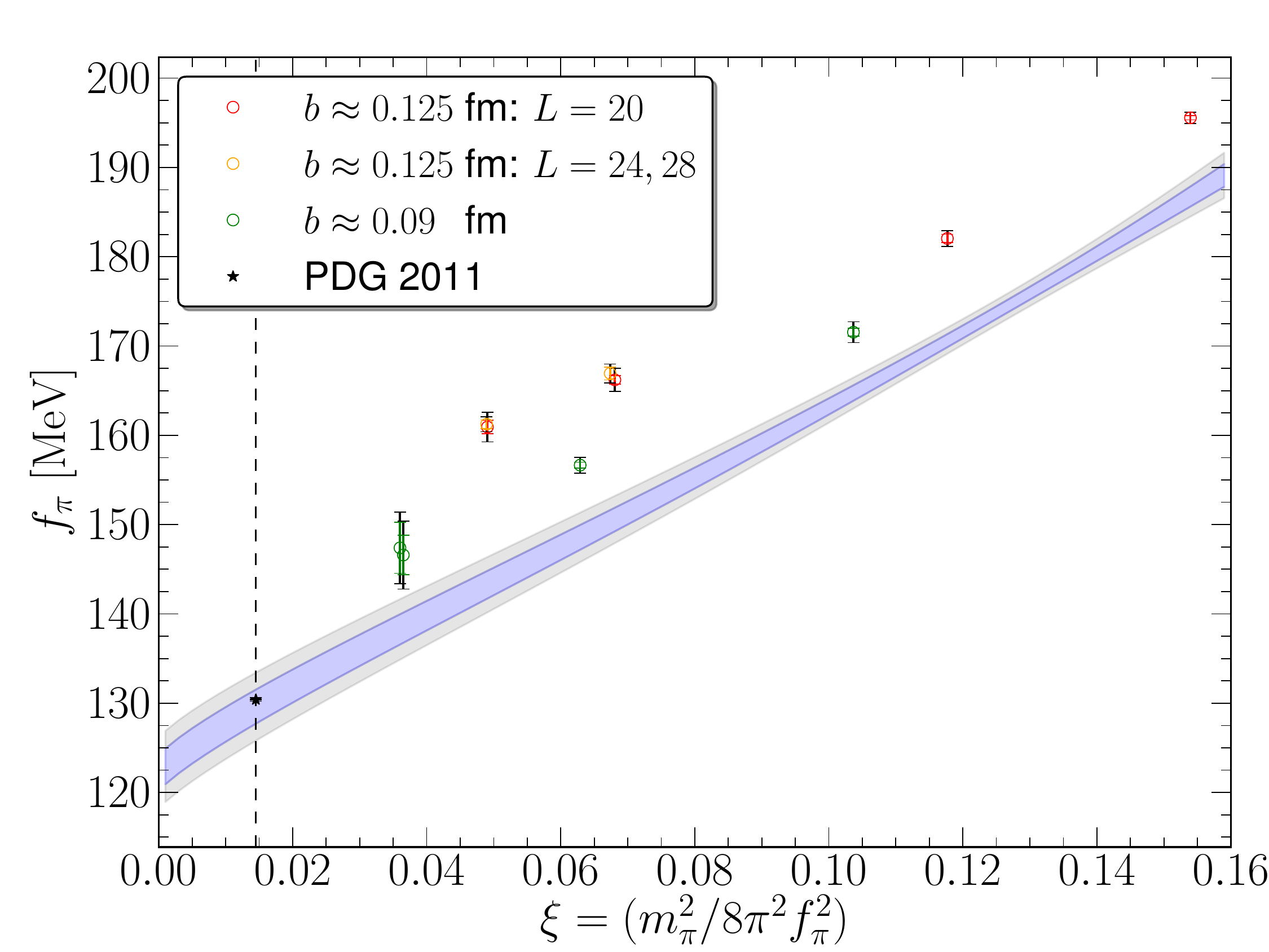}
\end{tabular}
\caption{\label{fig:MANNLO}
The result of NLO MA$\chi$PT plus NNLO
    $\chi$PT fit E described in the text, extrapolated to the infinite-volume and continuum limits.
The star denotes the experimentally determined value of $f_{\pi^+}$ (not
used in the fitting), listed in the Particle Data Group (PDG).
}
\end{figure}

Figure~\ref{fig:l3l4hist} shows Monte Carlo histograms of the extracted  values of
$\bar{l}_3$ and $\bar{l}_4$ using the Q weights to 
determine the ratio of samples to draw from each of fits A-E.
The result of fit E for $f_\pi$, extrapolated to the infinite-volume and continuum limits is
displayed in Fig.~\ref{fig:MANNLO}.
The inner (colored) band represents the 68\% statistical confidence interval
while the outer (gray) band results from the 68\% statistical and systematic
uncertainties combined in quadrature.
The dashed vertical line is located at $\xi^\textrm{phy}$ determined from
Eq.~\eqref{eq:mpifpi_phys}.
%
%

%
%
\subsection{Convergence of the $SU(2)$ Chiral Expansion \label{sec:ChiPT_converge} }

\begin{figure}[ht]
\begin{tabular}{cc}
\includegraphics[width=0.49\textwidth]{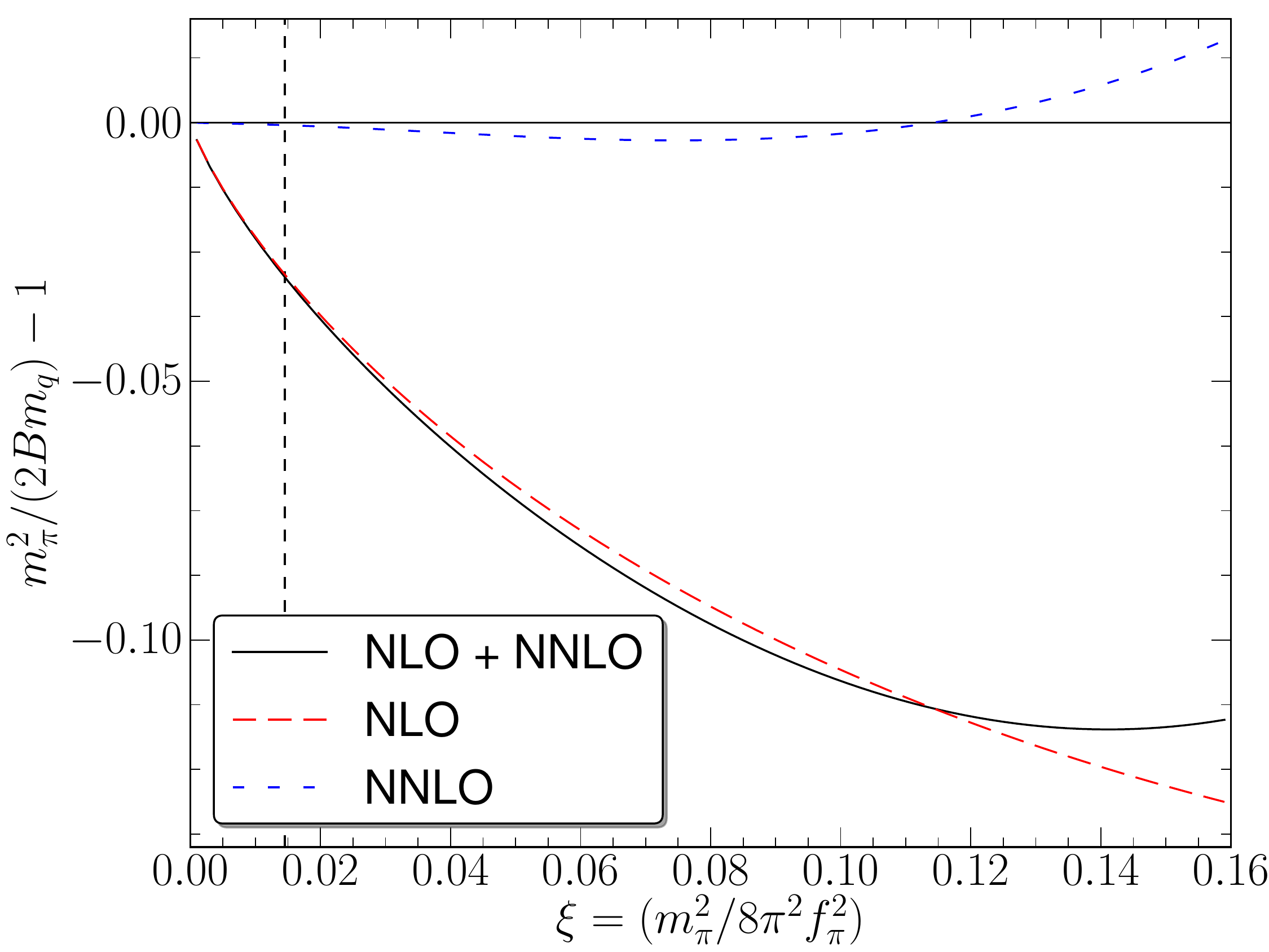}
&
\includegraphics[width=0.49\textwidth]{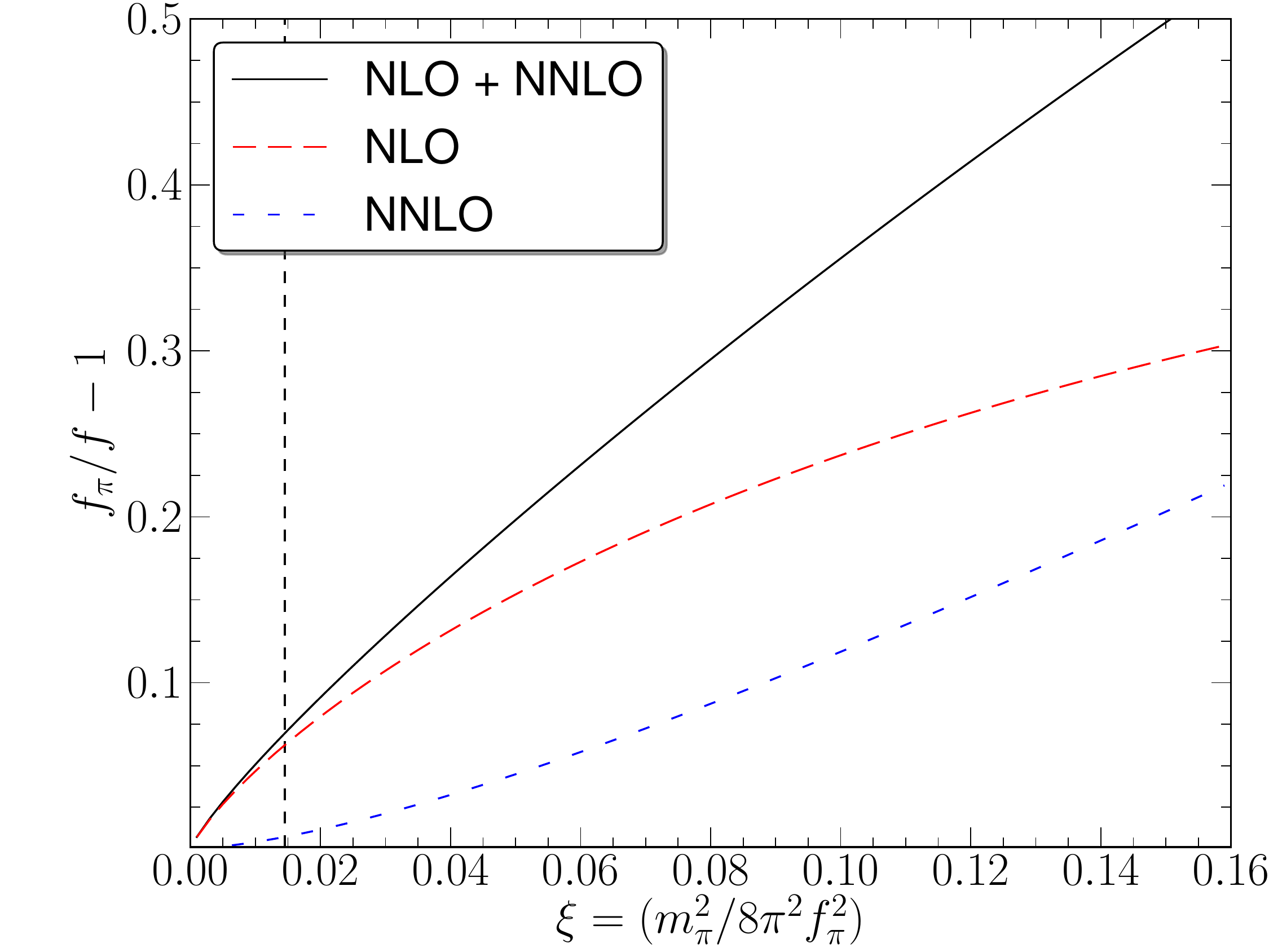}
\end{tabular}
\caption{\label{fig:mpifpi_expansion}
The NLO and NNLO contributions to $\frac{m_\pi^2}{2 B m_q} - 1 $ (left panel)
and 
$\frac{f_\pi}{f} - 1$ (right panel).  Both of these quantities vanish in the
chiral-limit.
The larger (red) dashed curves are the NLO contributions and the smaller (blue)
dashed curves 
are the NNLO contributions.  
The solid (black) curve is the entire NLO + NNLO value.
}
\end{figure}

\noindent
With the analyses performed in the previous section in hand, 
the  convergence of the two-flavor chiral expansion can be explored.
The resulting NLO and NNLO contributions to the quantities
\begin{align}
&\frac{m_\pi^2}{2 B m_q} - 1&
&\textrm{and}&
&\frac{f_\pi}{f} - 1\, ,&
\end{align}
(both of which vanish in the chiral-limit) are shown in Fig.~\ref{fig:mpifpi_expansion}.
%
%
In both cases (the left and right panels of Fig.~\ref{fig:mpifpi_expansion}), 
it is the continuum limit and infinite-volume limit extrapolations that are displayed.
In the case of $m_\pi$, the NNLO contributions are negligible over  most of the
range of $\xi$ used in our fits.
Further, the total corrections to $m_\pi$ are small, being less than $\sim 15\%$ 
over  the full range of quark-masses.
In contrast, the corrections to $f_\pi$ become substantial at the heavier pion
masses,   exceeding $\sim 50\%$ at the heaviest mass considered.  
Further, at the modest value of $\xi \gsim 0.08$  the NNLO corrections
become significant compared to the NLO corrections.

\begin{figure}
\begin{tabular}{cc}
\includegraphics[width=0.49\textwidth]{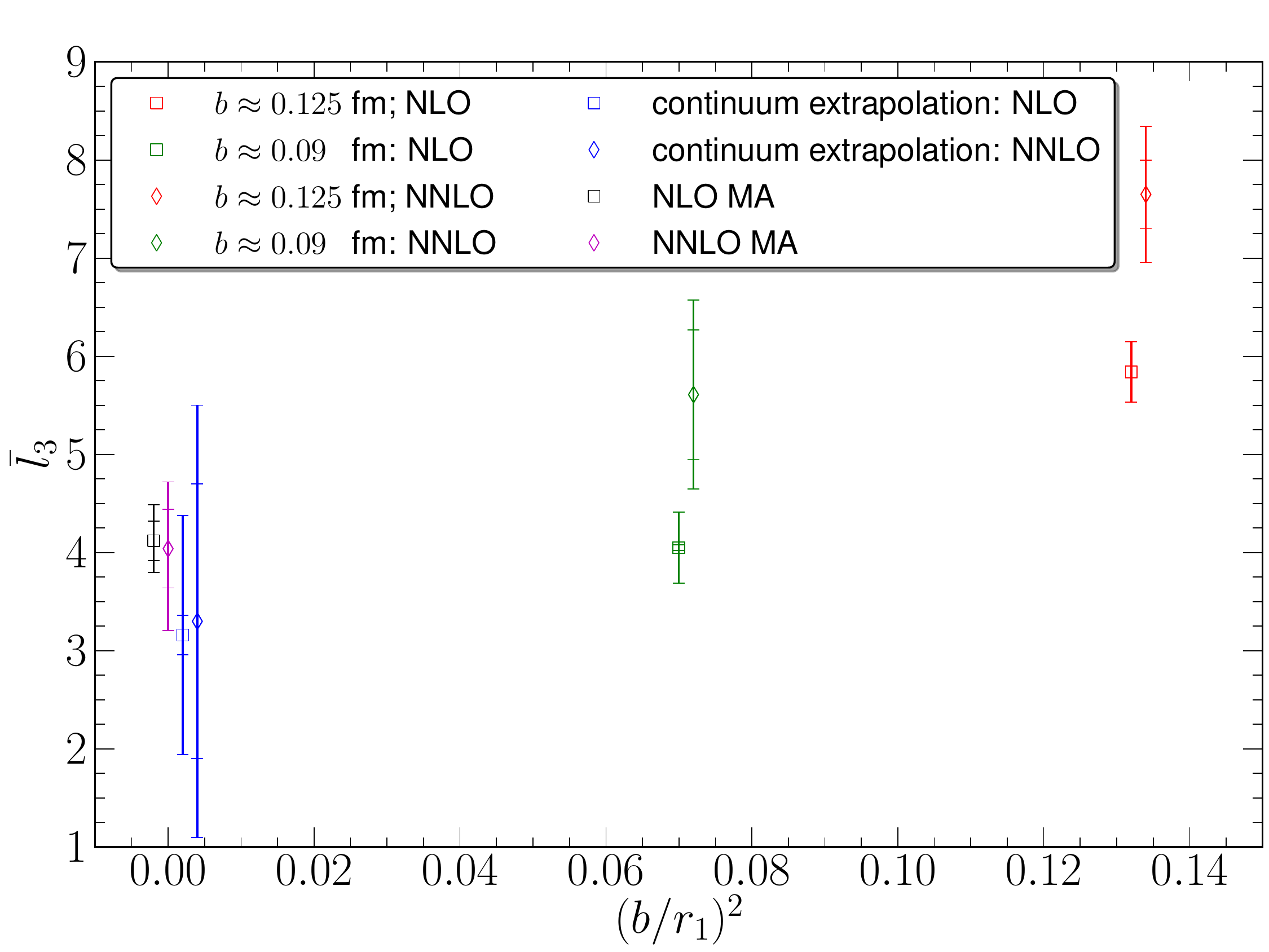}
&
\includegraphics[width=0.49\textwidth]{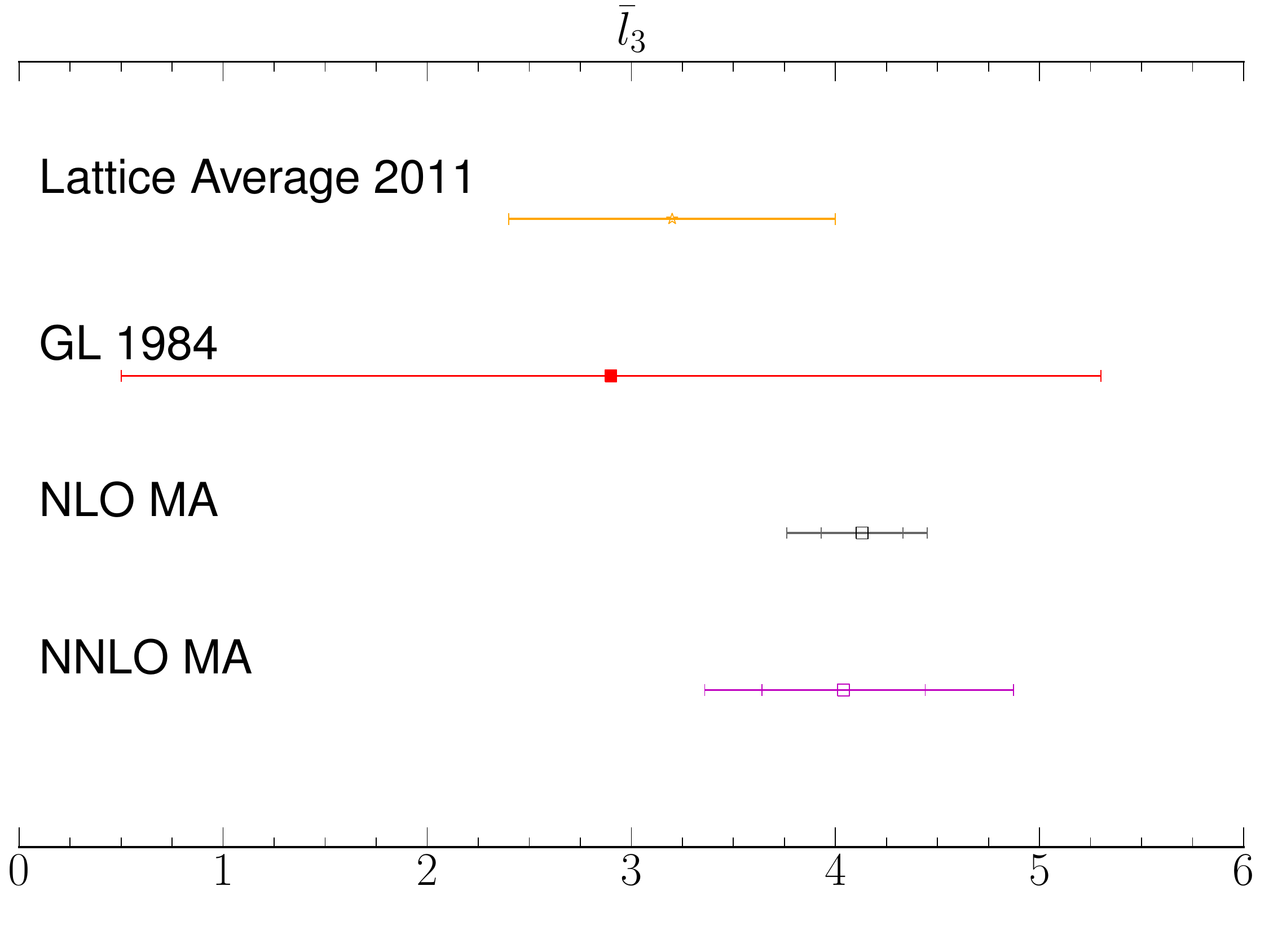}
\end{tabular}
\caption{\label{fig:l3}
The present determination of $\bar{l}_3$ (left panel),
and its comparison to the Lattice QCD average value~\cite{Colangelo:2010et} and
phenomenological results (right panel).
Some of the $\bar{l}_3$ results in the left panel
have been given small offsets in  $(b/r_1)^2$
for presentations reasons.
}
\end{figure}
In the left panel of Fig.~\ref{fig:l3}, the determination of
$\bar{l}_3$ is shown.  The results of the fixed lattice-spacing $\chi$PT analysis
from Sec.~\ref{sec:ChiPT_Vanilla_NNLO} is displayed, as well as the
continuum extrapolated value.  Also shown are the values extracted
from MA$\chi$PT at NLO, and from NLO MA$\chi$PT supplemented with
continuum NNLO $\chi$PT, as discussed in Sec.~\ref{sec:ChiPT_MA_NLO}
and Sec.~\ref{sec:ChiPT_MA_NNLO}, respectively.  The results of the
MA$\chi$PT analyses are consistent with the continuum extrapolated
results, but with smaller uncertainties.  This is not surprising as
the mixed-action framework allows a simultaneous treatment of
calculational results from multiple lattice spacings.  This
consistency lends confidence in the entire analysis.  In the right panel
of Fig.~\ref{fig:l3}, the extraction is compared to the original
estimates by Gasser and Leutwyler~\cite{Gasser:1983yg} as well as to
the recent Lattice QCD average~\cite{Colangelo:2010et}.
%
\begin{figure}
\begin{tabular}{cc}
\includegraphics[width=0.49\textwidth]{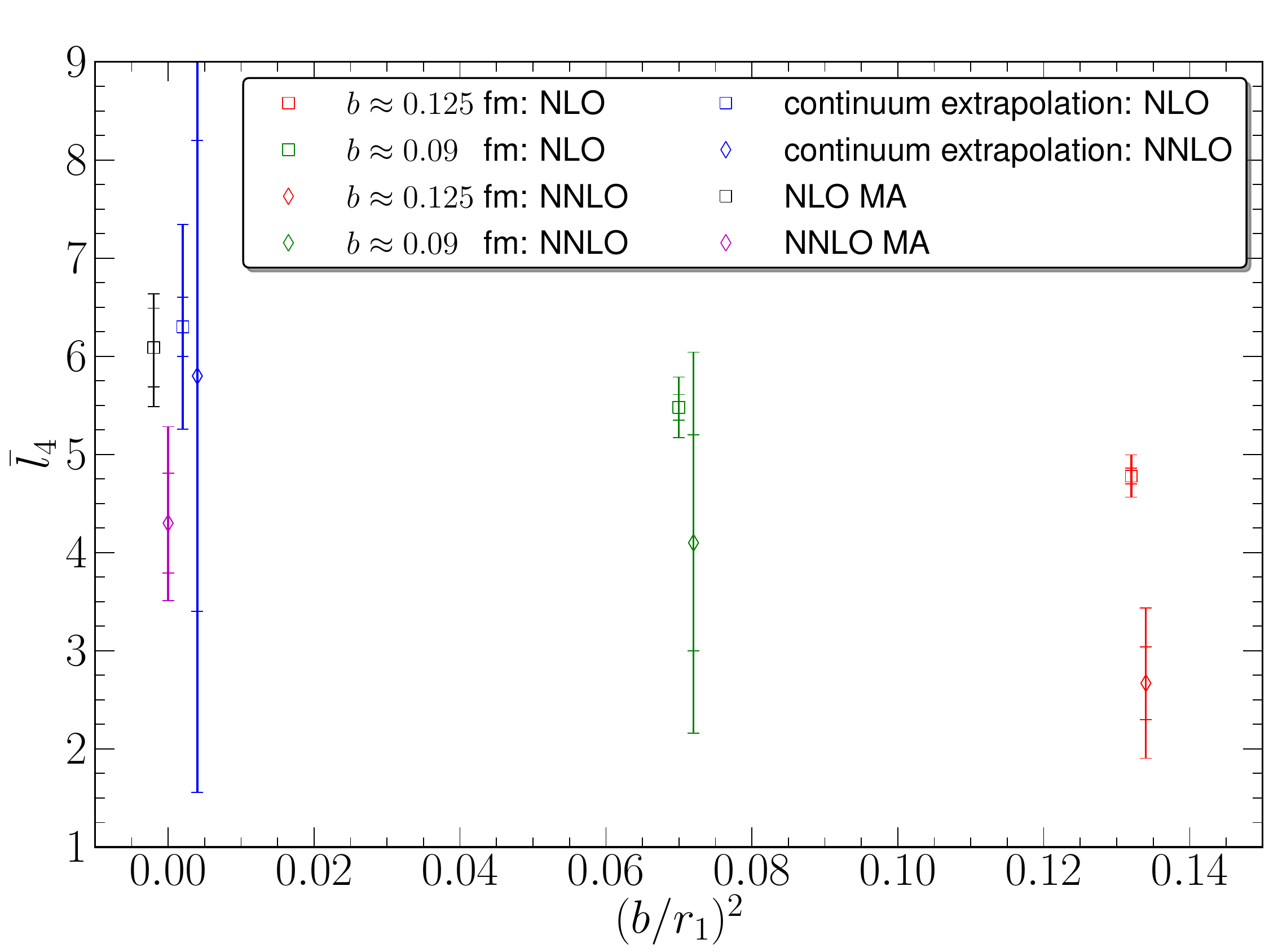}
&
\includegraphics[width=0.49\textwidth]{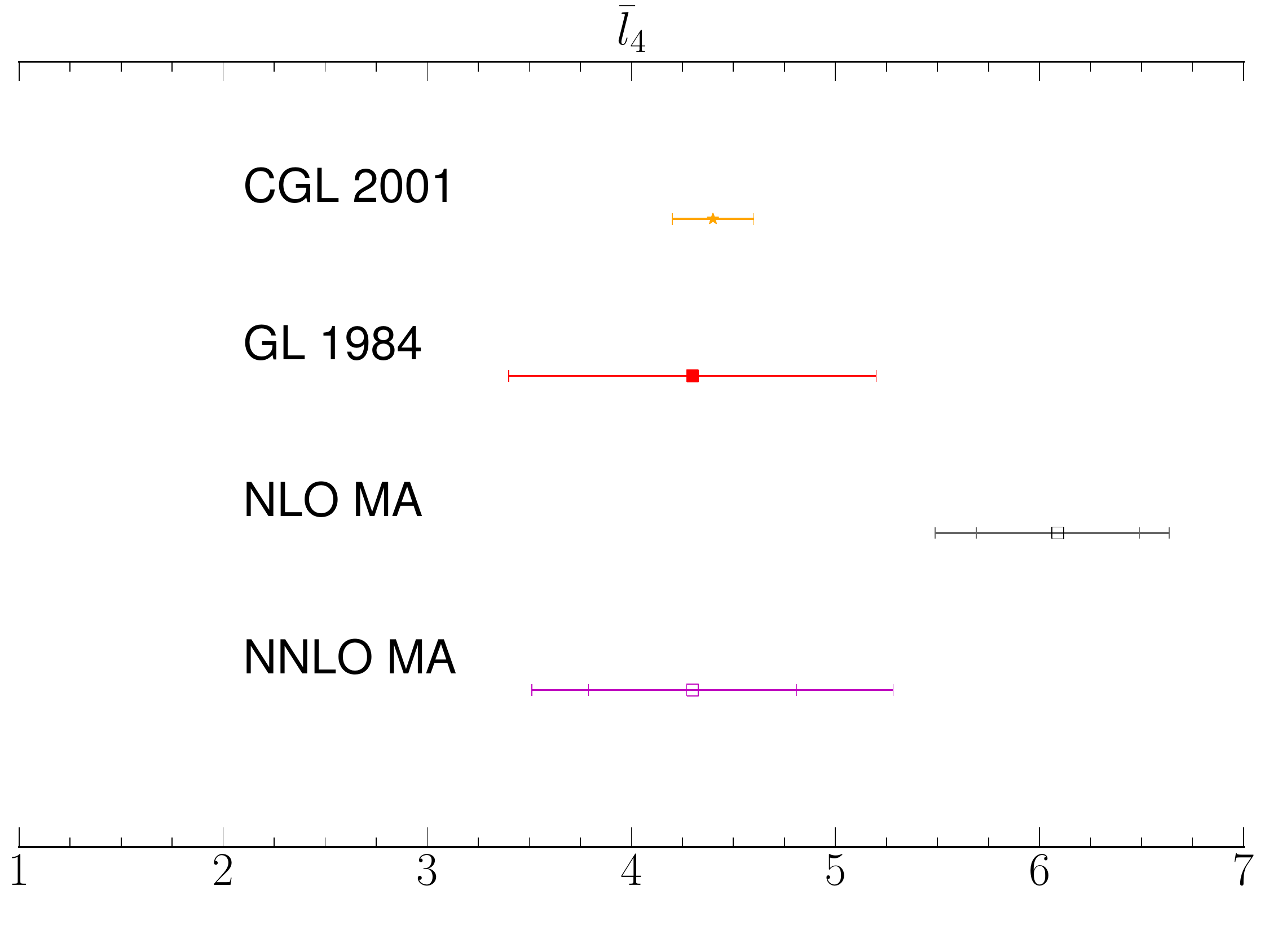}
\end{tabular}
\caption{\label{fig:l4}
The present determination of $\bar{l}_4$ (left panel), 
and its comparison with phenomenological results (right panel).  
(Ref.~\cite{Colangelo:2010et} does not currently provide a Lattice QCD average
value for this quantity.)
Some of the $\bar{l}_4$ results in the left panel
have been given small offsets in  $(b/r_1)^2$
for presentations reasons.  CGL 2001 refers to Ref.~\cite{Colangelo:2001df}.
}
\end{figure}
In Fig.~\ref{fig:l4}, 
the analogous results for $\bar{l}_4$ are displayed, although
Ref.~\cite{Colangelo:2010et} does not provide an average value 
(citing insufficient reporting of the associated systematic uncertainties).

%
%
\section{Results and Discussion \label{sec:Results} }

\noindent
We have performed precision calculations of the pion mass and the pion decay
constant with mixed-action Lattice QCD.  
Calculations using domain-wall valence
quarks and staggered sea-quarks were performed on a number of ensembles of MILC
gauge-field configurations at different light-quark-masses, two
lattice spacings, different volumes and different extents of the fifth
dimension. 
Using the two lattice spacings and the multiple light-quark-masses, the results of these calculations were extrapolated to the continuum, to infinite-volume and to the physical pion mass.
Ideally, continuum extrapolations would be performed with more than two lattice spacings.
While this is not possible with the present numerical results, the two methods used to quantify uncertainties associated with the continuum extrapolation from the two lattice spacings used in this work are found to give the same results within uncertainties.
One method involved using two-flavor $\chi$PT to extract the LECs, which
implicitly include lattice-spacing artifacts.  LECs calculated at two
different lattice spacings were then extrapolated to the continuum.
It is found that NLO $\chi$PT fails to describe the results of the Lattice
calculations of $m_\pi$, while NNLO $\chi$PT appears to be consistent with them.
The second method was to use MA$\chi$PT where the lattice-spacing artifacts
are explicit, and the extracted LECs are those of the continuum, up to higher
order contributions.  A hybrid analysis was motivated to be sufficient, where
the mixed-action NLO contributions were combined with continuum NNLO
contributions to provide reliable extractions of the LECs.
These analyses have provided determinations of the Gasser-Leutwyler
coefficients $\bar{l}_3$ and $\bar{l}_4$, 
\begin{align}
&\bar{l}_3 =  \lthree&
&\textrm{and}&
&\bar{l}_4 = \lfour&
\end{align}
These values are consistent with the (lattice) averaged values reported in Ref.~\cite{Colangelo:2010et}. 
Our analysis also provides
\begin{equation}
\frac{f_\pi}{f} = \fpif\, ,
\end{equation}
which is to be compared to the lattice averaged  value of $f_\pi / f = 1.073(15)$.
Combined with the experimental value for $f_\pi^{phy} = 130.4$~MeV, a value of $f = 122.8(3.0)({}^{4.6}_{4.8})$~MeV is found (we have not accounted for explicit isospin breaking effects, but these are expected to be small).
In Table~\ref{tab:Compare}, the present results are compared with those of the most recent calculations from other lattice collaborations.
Further, the extrapolated value of $r_1 f_\pi$ and the experimentally measured
value  of $f_{\pi^+}$ provides a determination of the physical scale $r_1$, 
\begin{equation}
r_1 = \rone  \textrm{ fm}\, ,
\end{equation}
which is to be compared with the MILC determination (on the same ensembles) 
of $r_1 = 0.311(2)({}^3_8)$~fm.
It is interesting to note that, despite greatly enhanced statistics on the same
ensembles of MILC gauge-field configurations,
the uncertainty that we have obtained in the calculation of 
$f_\pi$ is somewhat larger than that obtained in Ref.~\cite{Aubin:2008ie}.

\begin{table}
\caption{\label{tab:Compare}
Comparison with most recent results from various lattice collaborations.}
\begin{ruledtabular}
\begin{tabular}{llllll}
Collaboration& Reference & $N_f$& $f_\pi / f$& $\bar{l}_3$& $\bar{l}_4$\\
\hline
MILC 10 [$SU(3)$]& \cite{Bazavov:2010hj}& $2+1$& 1.06(5)& 3.18(50)(89)& 4.29(21)(82)\\
MILC 10A [$SU(2)$]& \cite{Bazavov:2010yq}& $2+1$& 1.05(1)& $2.85(81)({}^{37}_{92})$& $3.98(32)({}^{51}_{28})$ \\
RBC/UKQCD 10A& \cite{Aoki:2010dy}& $2+1$& --& 2.57(18)& 3.83(09) \\
ETM 10& \cite{Baron:2010bv}& $2+1+1$& 1.076(2)(2)& 3.70(07)(26)& 4.67(03)(10)\\
ETM 09C& \cite{Baron:2009wt}& 2& $1.0755(6)({}^{08}_{94})$& $3.50(9)({}^{09}_{30})$& $4.66(4)({}^{04}_{33})$\\
PACS-CS 08 [$SU(3)$]& \cite{Aoki:2008sm}& $2+1$& 1.062(8)& 3.47(11)& 4.21(11) \\
PACS-CS 08 [$SU(2)$]& \cite{Aoki:2008sm}& $2+1$& 1.060(7)& 3.14(23)& 4.04(19) \\
JLQCD/TWQCD 08A& \cite{Noaki:2008iy}& 2& 1.17(4)& $3.38(40)(24)({}^{31}_0)$& $4.12(35)(30)({}^{31}_0)$\\
RBC/UKQCD 08& \cite{Allton:2008pn}& $2+1$& 1.080(8)& 3.13(33)(24)& 4.43(14)(77) \\
\hline
FLAG Avg.& \cite{Colangelo:2010et}& --& 1.073(15)& 3.2(8)& -- \\
\hline
NPLQCD [this work]& & $2+1$& $\fpif$& $\lthree$& $\lfour$ \\
\end{tabular}
\end{ruledtabular}
\end{table}

The systematics in the calculations arising from the finite lattice volume and
from residual chiral symmetry breaking due to the finite fifth-dimensional
extent of the domain-wall action have been explored and quantified.
Previously, residual chiral symmetry breaking contributions were identified 
to be the dominant source of uncertainty in Lattice QCD 
predictions of the $I=2\ \pi\pi$ scattering length~\cite{Beane:2007xs}.
While the present analysis has not been able to precisely determine these
effects,  the  analysis resulted in constraints on the size of these
contributions,
\begin{align}
&\bar{l}_3^\textrm{res} = \lthreeRes\, ,&
&\bar{l}_4^\textrm{res} = \lfourRes\, ,&
\end{align}
which in turn can be used to reduce the uncertainties in the $I=2\ \pi\pi$ scattering length predictions.

The predicted NLO mixed-action finite-volume contributions to the pion mass appear to be
incompatible with the results of the Lattice QCD calculations,
suggesting  the importance of higher orders in the MA$\chi$PT expansion.  
A 30\% systematic uncertainty is assigned to the NLO finite-volume contributions to account
for NNLO effects, leading to a consistent description of the results.

In Table~\ref{tab:ErrorBudget} the contributions to the total uncertainty from the various systematics are displayed.
While the discretization and residual chiral symmetry breaking effects have some impact on the determination of the LECs, it is clear from this summary table that the dominant uncertainty is due to the chiral extrapolation.  Having further numerical results at lighter pion masses is the single most important systematic to address to improve upon the present work.

In conclusion, we have found that a careful two-flavor low-energy effective
field theory analysis of the Lattice QCD calculations of the pion mass and its
decay constant can reliably determine the NLO Gasser-Leutwyler
coefficients,  $\bar{l}_3$ and $\bar{l}_4$, which are found to be in good
agreement with the average of other determinations. 
In particular, mixed-action chiral perturbation theory which includes
lattice-spacing artifacts explicitly, provides a reliable framework with which
to perform chiral extrapolations of $m_\pi$ and $f_\pi$ to the physical
light quark-masses, and to determine $\bar{l}_3$ and $\bar{l}_4$.

\begin{table}
\caption{\label{tab:ErrorBudget}
Error budget for current work expressed as relative uncertainties.
}
\begin{ruledtabular}
\begin{tabular}{c|c|ccccccc}
Quantity& Total & Statistical & Chiral & Continuum& Volume& $m^\textrm{res}$ & $m_s^\textrm{tune}$\\
& uncertainty& uncertainty& extrapolation& extrapolation& extrapolation& \\
\hline
$\bar{l}_3$& 19\%& 10\%& 15\%& 5\%& 0\%& 2.7\%& 0\% \\
$\bar{l}_4$& 21\%& 7\%& 19\%& 4\%& 0\%& 4\% & 0\% \\
$f_\pi / f$& 4.6\%& 2.4\% & 3.9\%& 0\%& 0\%& 0\%& 0\%
\end{tabular}
\end{ruledtabular}
\end{table}

\acknowledgments

\noindent We would to thank the LHP Collaboration for their light quark
propagators computed on the \coarse MILC ensembles.  
We thank C.~DeTar for help with the HMC generation of the large volume m007m050 ensemble.
We thank C.~Bernard for providing the updated values of $r_1$ from 
MILC as well as those extrapolated to the physical values of the light quark-masses.  
We thank G.~Colangelo for valuable conversations and R.~Edwards and B.~Joo for
developing \texttt{qdp++} and \texttt{chroma}~\cite{Edwards:2004sx}.  
We would also like to thank H.-W. Lin for comments on the manuscript.
We acknowledge computational support from the USQCD SciDAC project, 
National Energy Research Scientific Computing Center
(NERSC, Office of Science of the DOE, Grant No.~DE-AC02-05CH11231), 
the UW
HYAK facility, Centro Nacional de Supercomputaci\'on (Barcelona,
Spain), LLNL, the Argonne Leadership Computing Facility at Argonne
National Laboratory (Office of Science of the DOE, under contract
No.~DE-AC02-06CH11357), and the NSF through Teragrid resources
provided by TACC and NICS under Grant No.~TG-MCA06N025.  
SRB was supported in part by the NSF CAREER Grant No. PHY-0645570. The Albert Einstein Center for Fundamental Physics is supported by the Innovations- und Kooperationsprojekt C-13 of the Schweizerische Universit\"{a}tskonferenz SUK/CRUS.
The work of AP is supported
by the contract FIS2008-01661 from MEC (Spain) and FEDER and from the RTN Flavianet MRTN-CT-2006-035482 (EU).
MJS is supported in
part by the DOE Grant No.~DE-FG03-97ER4014.  
WD and KO were supported
in part by DOE Grants No.~DE-AC05-06OR23177 (JSA) and
No.~DE-FG02-04ER41302.  
WD was also supported by DOE OJI Grant
No.~DE-SC0001784 and Jeffress Memorial Trust, Grant No.~J-968.  
KO was
also supported in part by NSF Grant No.~CCF-0728915 and DOE OJI Grant
No.~DE-FG02-07ER41527.  
AT was supported by NSF Grant No.~PHY-0555234
and DOE Grant No.~DE-FC02-06ER41443.  
The work of TL was performed
under the auspices of the U.S.~Department of Energy by LLNL under
Contract No.~DE-AC52-07NA27344.  
The work of AWL was supported in part by the
Director, Office of Energy Research, Office of High Energy and Nuclear
Physics, Divisions of Nuclear Physics, of the U.S. DOE under Contract
No.~DE-AC02-05CH11231.

\bibliography{SU2_MA}

\end{document}